\documentclass[floatfix,showpacs,prb,aps,a4paper,twocolumn]{revtex4}

 \usepackage{bm}
 \usepackage{amsmath}
 \usepackage{amssymb}
 \usepackage{latexsym}
 \usepackage{amsfonts}
 \usepackage{epsfig}
 \usepackage{subfigure}
 \usepackage{color}
 \definecolor{darkblue}{rgb}{0,0,.5}
 \usepackage[linktocpage, colorlinks=true ,linkcolor=darkblue, citecolor=darkblue]{hyperref}
 \usepackage[all]{hypcap}

\usepackage{bbold}
\usepackage{upgreek}
\usepackage{dsfont}

\newcommand{\ket}[1]{\left|#1\right>}
\newcommand{\bra}[1]{\left<#1\right|}

\newcommand{\expval}[1]{{\left<#1\right>}}

\newcommand{\com}[1]{{\left[#1\right]}}

\newcommand{\nn}{\nonumber\\}

\newcommand{\f}[1]{\mbox{\boldmath$#1$}}
\newcommand{\abs}[1]{\left| #1 \right|}

\newcommand{\HS}{\mathcal{H}_{\rm S}}

\newcommand{\HB}{\mathcal{H}_{\rm B}}
\newcommand{\HSB}{\mathcal{H}_{\rm SB}}

\newcommand{\RS}{\rho_{\rm S}}
\newcommand{\RB}{\rho_{\rm B}}

\newcommand{\fRS}{\boldsymbol \rho_{\rm S}}

\newcommand{\fR}{\boldsymbol \rho}

\newcommand{\ii}{\mathrm{i}}
\newcommand{\Id}[1]{\mathds{1}}

\newcommand{\eL}{\upvarepsilon_{\rm L}}
\newcommand{\eR}{\upvarepsilon_{\rm R}}
\newcommand{\eO}{\upvarepsilon_0}
\newcommand{\eP}{\upvarepsilon_+}
\newcommand{\eM}{\upvarepsilon_-}
\newcommand{\ePM}{\upvarepsilon_2}
\newcommand{\etL}{\bar\upvarepsilon_{\rm L}}
\newcommand{\etR}{\bar\upvarepsilon_{\rm R}}

\newcommand{\ekR}{\upvarepsilon_{k{\rm R}}}
\newcommand{\ekL}{\upvarepsilon_{k{\rm L}}}
\newcommand{\dL}{d_{\rm L}}
\newcommand{\dR}{d_{\rm R}}

\newcommand{\aq}{a_{q}}
\newcommand{\tkL}{t_{k{\rm L}}}
\newcommand{\tkR}{t_{k{\rm R}}}

\newcommand{\hqL}{h_{q{\rm L}}}
\newcommand{\hqR}{h_{q{\rm R}}}
\newcommand{\hqa}{h_{q\sigma}}

\newcommand{\tc}{T_{\rm c}}
\newcommand{\ttc}{{\bar T}_{\rm c}}

\newcommand{\dE}{\Delta \upvarepsilon}
\newcommand{\bL}{\beta_{\rm L}}
\newcommand{\bR}{\beta_{\rm R}}
\newcommand{\bPH}{\beta_{\rm ph}}
\newcommand{\mL}{\mu_{\rm L}}
\newcommand{\mR}{\mu_{\rm R}}
\newcommand{\gL}{\Gamma_{\rm L}}
\newcommand{\gR}{\Gamma_{\rm R}}
\newcommand{\wc}{\omega_{\rm c}}
\newcommand{\nB}{n_{\rm B}}

\begin{document}

%%%%%%%%%%%%%%%%%%%%%%%%%%%%%%%%%%%%%%%%%%%%%%%%%%%%%%%%%%%%%%%%%%%%%%%%%%%%%%%
%%%%%%%%%%%%%%%%%%%%%%%%%%%%%%%%%%%%%%%%%%%%%%%%%%%%%%%%%%%%%%%%%%%%%%%%%%%%%%%

\title{Thermodynamics of the polaron master equation at finite bias}
\author{Thilo Krause$^{1}$}\email{tkrause@physik.tu-berlin.de}
\author{Tobias Brandes$^1$}
\author{Massimiliano Esposito$^2$}
\author{Gernot Schaller$^1$}\email{gernot.schaller@tu-berlin.de}
\affiliation{$^1$ Institut f\"ur Theoretische Physik, Technische Universit\"at Berlin, Hardenbergstr. 36, D-10623 Berlin, Germany\\
$^2$ Complex Systems and Statistical Mechanics, University of Luxembourg, L-1511 Luxembourg, Luxembourg}

%%%%%%%%%%%%%%%%%%%%%%%%%%%%%%%%%%%%%%%%%%%%%%%%%%%%%%%%%%%%%%%%%%%%%%%%%%%%%%%
%%%%%%%%%%%%%%%%%%%%%%%%%%%%%%%%%%%%%%%%%%%%%%%%%%%%%%%%%%%%%%%%%%%%%%%%%%%%%%%

\begin{abstract}
We study coherent transport through a double quantum dot.
Its two electronic leads induce electronic matter and energy transport and a phonon reservoir contributes
further energy exchanges.
By treating the system-lead couplings perturbatively, whereas the
coupling to vibrations is treated non-perturbatively in a polaron-transformed frame, 
we derive a thermodynamic consistent low-dimensional master equation.
When the number of phonon modes is finite, a Markovian description is only possible 
when these couple symmetrically to both quantum dots.
For a continuum of phonon modes however, also asymmetric couplings can be described
with a Markovian master equation.
We compute the electronic current and dephasing rate.
The electronic current enables transport spectroscopy of the phonon frequency and
displays signatures of Franck-Condon blockade.
For infinite external bias but finite tunneling bandwidths, we find oscillations in the 
current as a function of the internal bias due to the electron-phonon coupling.
Furthermore, we derive the full fluctuation theorem and show its identity to the 
entropy production in the system.
\end{abstract}

%%%%%%%%%%%%%%%%%%%%%%%%%%%%%%%%%%%%%%%%%%%%%%%%%%%%%%%%%%%%%%%%%%%%%%%%%%%%%%%
%%%%%%%%%%%%%%%%%%%%%%%%%%%%%%%%%%%%%%%%%%%%%%%%%%%%%%%%%%%%%%%%%%%%%%%%%%%%%%%


%%%%%%%%%%%%%%%%%%%%%%%%%%%%%%%%%%%%%%%%%%%%%%%%%%%%%%%%%%%%%%%%%%%%%%%%%%%%%%%
%%%%%%%%%%%%%%%%%%%%%%%%%%%%%%%%%%%%%%%%%%%%%%%%%%%%%%%%%%%%%%%%%%%%%%%%%%%%%%%

\maketitle

%%%%%%%%%%%%%%%%%%%%%%%%%%%%%%%%%%%%%%%%%%%%%%%%%%%%%%%%%%%%%%%%%%%%%%%%%%%%%%%
%%%%%%%%%%%%%%%%%%%%%%%%%%%%%%%%%%%%%%%%%%%%%%%%%%%%%%%%%%%%%%%%%%%%%%%%%%%%%%%
Electronic transport through low-dimensional systems, e.g.\ quantum dots or molecular junctions, has been a vivid research field over the last years.
In part, this has been triggered by the fact that single molecules or quantum dot configurations are promising candidates for a variety of applications 
such as e.g.\ charge~\cite{gorman2005} and spin~\cite{trauzettel2007} qubits or single photon emitters which are for example realized in semiconductor nanowires~\cite{claudon2010,holmes2014}.
%emitters that are realized in semiconductor nanowires~\cite{claudon2010,holmes2014}.
%
For efficient device performance a detailed understanding of electronic interplay with its environment, e.g.\ optical modes~\cite{liu2014} is important.
In particular, the interaction with vibrational modes has been studied in order to 
reveal quantum phenomena such as additional decoherence~\cite{ballmann2012,hartle2013,gamble2012,kaer2014}.
%reveal quantum phenomena by phonon spectroscopy~\cite{brandes1999,ueda2010,hartle2011,roulleau2011
%,franke2012}.
%
%These may lead to additional decoherence~\cite{ballmann2012,hartle2013,gamble2012,kaer2014} but also
Moreover, phonon spectroscopy~\cite{brandes1999,ueda2010,roulleau2011,franke2012} can be used to visualize quantum effects
%manifest 
in transport characteristics such as Franck-Condon blockade and giant 
Fano factors~\cite{koch2005,koch2006,leturcq2009,donabidowicz2012,huetzen2012a,santamore2013,timm2013,koch2014}.
It was also proposed to use bias-controlled electronic transport to selectively excite vibrational modes~\cite{volkovich2011a}.
Furthermore, also from a more classical perspective, the study of thermo-electric effects in phonon-coupled 
nanojunctions~\cite{segal2005% probe thermoelectric effects with current
,galperin2008%influence of molecular vibration on the Seebeck coefficient
,dubi2009%general
,dubi2011%review
,hartle2011%vibrational instabilities are related to electron-hole pair creation processes
} 
-- e.g.\ the conversion of heat to work~\cite{rutten2009%efficiency conversion
,entin_wohlman2010a%conversion heat and work
,sanchez2013%charge current can be generated by heat conversio
,sothmann2015%review: heat work conversion
} 
or local cooling~\cite{esposito2009,%efficiency for cooling a system
galperin2009a,%current-induced cooling
arrachea2014a} -- 
leads to interesting new questions.

A crucial parameter for understanding many of these effects is the coupling strength between electronic transport and phonon modes.
Therefore, weak~\cite{haupt2009,park2011} and strong~\cite{galperin2006,nicolin2011a,nicolin2011b,ren2011,Tkoch2014} coupling regimes have 
been studied with different methods.  
Recently, the possibility to tune the coupling strength in the experiment has been suggested~\cite{benyamini2014}.  

Even though most authors consider simplified models consisting of a few-level-system 
coupled to either a discrete (Anderson-Holstein model)~\cite{holstein1959,anderson1961,koch2006a,galperin2007a} or multiple~\cite{schinabeck2014} phonon modes, 
the inclusion of the vibration dynamics in the system leads to an infinite-dimensional Hilbert space and therefore large computational complexity.
This problem has triggered the development of many different methods, such as master equations~\cite{hartle2011,kast2011a}, Full Counting Statistics approaches~\cite{avriller2009,urban2010,wang2012}, Lindblad kinetic equations~\cite{dzhioev2012a}, Greens functions~\cite{galperin2006,hartle2009}, 
or higher-order methods~\cite{leijnse2008a}.
For time-dependent studies often the multilayer multiconfiguration time-dependent Hartree method~\cite{wang2009a,albrecht2012,wilner2014a} or real-time path integral approaches are applied~\cite{muehlbacher2008a,schiro2009,simine2013a}.
Additional complexity arises if intrinsic tunneling between multiple system states and their coupling to phonon modes is taken into account.
Among the conceptually simple master equation approaches, the polaron-transformed master equation~\cite{weiler2012} often allows a simple 
diagonalization of the system Hamiltonian, which has renewed interest in the phonon master equation in the finite-bias regime~\cite{maier2011,walter2013a}.
As long as the phonons are contained within the system, this leads to thermodynamic consistency but does not solve the curse of dimensionality.

When, in contrast, within a strong electron-phonon-coupling scenario the vibrations are treated as part of a reservoir, 
thermodynamic consistency is non-trivial even for a single electronic level~\cite{schaller2013a}.
Here, the proof of the fluctuation theorem~\cite{crooks1999} offers a well known tool 
%for proving thermodynamic consistency and for understanding the thermodynamic properties 
%of a model 
because it directly confirms the second law of thermodynamics\cite{seifert2005,esposito2007,esposito2010a}.
In particular, in this paper we discuss the derivation of a phonon master equation for a double quantum dot model, introduced in Sec.~\ref{sec:Model}, coupled to macroscopic electronic leads and either a discrete or continuous phonon reservoir. 
Even in absence of phonons, we explicitly allow for electronic tunneling within the system. 
We treat the phonons as part of a non-standard reservoir yielding a finite system Hilbert space and, 
thus, a low dimensional master equation with minimized computational requirements making the method applicable for the study of even larger systems. 
We put emphasis on the polaron transformation and its effect on the model Hamiltonian in terms of thermodynamic consistency. 
Staying in the polaron picture, we present a detailed derivation of the quantum master equation, see Sec.~\ref{sec:MasterEquation}, and prove its thermodynamic consistency by deriving the fluctuation theorem in Sec.~\ref{sec:FCSsingleMode}. 
Finally, in Sec.~\ref{sec:Results}, we analyze electronic current and dephasing rate for particular physical situations showing a Franck-Condon-like 
suppression in both quantities.
We also investigate the possibility of phonon spectroscopy experiments.
In addition, we discuss the performance of the model system as a thermoelectric generator converting
a temperature gradient into useful power.

%urban2010%FCS strong weak coupling

%%%%%%%%%%%%%%%%%%%%%%%%%%%%%%%%%%%%%%%%%%%%%%%%%%%%%%%%%%%%%%%%%%%%%%%%%%%%%%%
%%%%%%%%%%%%%%%%%%%%%%%%%%%%%%%%%%%%%%%%%%%%%%%%%%%%%%%%%%%%%%%%%%%%%%%%%%%%%%%
%%%%%%%%%%%%%%%%%%%%%%%%%%%%%%%%%%%%%%%%%%%%%%%%%%%%%%%%%%%%%%%%%%%%%%%%%%%%%%%
%%%%%%%%%%%%%%%%%%%%%%%%%%%%%%%%%%%%%%%%%%%%%%%%%%%%%%%%%%%%%%%%%%%%%%%%%%%%%%%

\section{Model}\label{sec:Model}

%%%%%%%%%%%%%%%%%%%%%%%%%%%%%%%%%%%%%%%%%%%%%%%%%%%%%%%%%%%%%%%%%%%%%%%%%%%%%%%
%%%%%%%%%%%%%%%%%%%%%%%%%%%%%%%%%%%%%%%%%%%%%%%%%%%%%%%%%%%%%%%%%%%%%%%%%%%%%%%

\subsection{Hamiltonian}\label{sec:Hamiltonian}

We consider a system made of a double quantum dot (DQD) in contact with multiple reservoirs $\mathcal{H}=\HS+\HB+\HSB$.
The reservoirs $\HB = \HB^{\rm el} + \HB^{\rm ph}$ and the system-bath coupling $\HSB = \HSB^{\rm el} + \HSB^{\rm ph}$ contain
electronic and phonon contributions, respectively.
The DQD Hamiltonian reads
\begin{eqnarray}
\label{eq:HS}
\HS&\equiv&\eL\dL^\dagger\dL+\eR\dR^\dagger\dR+T_{\rm c}(\dL\dR^\dagger+\dR\dL^\dagger)\nn
&&+U\dL^\dagger\dL\dR^\dagger\dR\,,
\end{eqnarray}
where $d_{\sigma}(d_{\sigma}^\dagger)$ annihilates (creates) an electron
in dot $\sigma$ with on-site energy $\upvarepsilon_{\sigma}$ ($\sigma \in
\{ \rm L,R \}$ throughout this paper), $T_{\rm c}$ is the internal electronic tunneling
amplitude, and $U$ is the Coulomb repulsion energy.
The system is connected to two electronic leads left and right held at thermal equilibrium
\begin{eqnarray}
\label{eq:HBel}
\HB^{\rm el}\equiv \sum_k \sum_{\sigma\in\{\rm L,R\}}
\upvarepsilon_{k,\sigma}c_{k,\sigma}^\dagger c_{k,\sigma}\,.
\end{eqnarray}
Here, the fermionic operator $c_{k,\sigma}(c_{k,\sigma}^\dagger)$
annihilates (creates) electrons in mode $k$ with energy
$\upvarepsilon_{k,\sigma}$.
Note that we do not distinguish between the electronic spins, which
implicitly assumes that e.g.\  the leads are completely polarized.
Electronic transport through the system is enabled by the dot-lead interaction Hamiltonian
\begin{eqnarray}
\label{eq:HSBel}
\HSB^{\rm el}\equiv\sum_{k,\sigma\in\{\rm L,R\}}(t_{k,\sigma} d_\sigma c_{k,\sigma}^\dagger+{\rm h.c.})\,,
\end{eqnarray}
with the tunneling amplitudes $t_{k,\sigma}$ (which we will treat perturbatively to second order later-on).

Additionally, the system is coupled to a bosonic reservoir
\begin{eqnarray}
\label{eq:HBph}
\HB^{\rm ph}&\equiv&\sum_q\omega_{q}\aq^\dagger\aq\,,
\end{eqnarray}
with phonon operator $a_q (a_q^\dagger)$ annihilating (creating) a phonon in mode $q$ with frequency $\omega_q$.
The electronic occupation of the system induces vibrations in the phonon bath via the
electron-phonon interaction Hamiltonian
\begin{eqnarray}
\label{eq:HSBph}
\HSB^{\rm ph}\equiv\sum_q \sum_{\sigma\in\{\rm L,R\}}(h_{q,\sigma}\aq+{\rm h.c.})d_\sigma^\dagger d_\sigma\,,
\end{eqnarray}
with the phononic absorption/emission amplitudes $h_{q,\sigma}$ (which we want to treat non-perturbatively later-on).

%%%%%%%%%%%%%%%%%%%%%%%%%%%%%%%%%%%%%%%%%%%%%%%%%%%%%%%%%%%%%%%%%%%%%%%%%%%%%%%
%%%%%%%%%%%%%%%%%%%%%%%%%%%%%%%%%%%%%%%%%%%%%%%%%%%%%%%%%%%%%%%%%%%%%%%%%%%%%%%

\subsection{Polaron transformation}\label{sec:PolaronTransformation}
In order to investigate the impact of strong electron-phonon coupling on electronic transport we perform the unitary Lang-Firzov (polaron)
transformation~\cite{mahan2000,brandes2005a}, $\bar{\mathcal{H}}=U \mathcal{H}U^\dagger$, with the unitary operator
$U=e^{\dL \dL^\dagger \mathcal{B}_{\rm L}+\dR \dR^\dagger \mathcal{B}_{\rm R}}$. 
The anti-hermitian operator $\mathcal{B}_{\sigma}$ is defined as 
\begin{eqnarray}
\label{eq:B_alpha}
\mathcal{B}_{\sigma}\equiv\sum_{q}(h_{q, \sigma}^\ast a_{q}^\dagger-h_{q,\sigma}a_{q})/\omega_{q}\,. 
\end{eqnarray}
The details of the polaron transformation are shown in Appendix~\ref{App:PolaronTrafo}.
After the polaron transformation, the Hamiltonian admits a new decomposition into system, interaction, 
and reservoir contributions.
It is important to note, however, that in general such decompositions are not unique:
For example, for a system Hamiltonian $H_{\rm S}$ and an interaction Hamiltonian of the general form $H_{\rm I} = \sum_\alpha A_\alpha \otimes B_\alpha$ with
system and reservoir operators $A_\alpha$ and $B_\alpha$, respectively, it is straightforward to see that
the transformation $H_{\rm S} \to H_{\rm S} + \sum_\alpha \kappa_\alpha A_\alpha$ and 
$H_{\rm I} \to \sum_\alpha A_\alpha \otimes \left(B_\alpha - \kappa_\alpha \f{1}\right)$ with numbers $\kappa_\alpha$
leaves the total Hamiltonian invariant.

We resolve this ambiguity by demanding that all thermal equilibrium expectation values of linear bath coupling operators should vanish.
We have observed that without imposing this requirement one would arrive at a thermodynamic inconsistent master equations (e.g.\ predicting non-vanishing currents at
global equilibrium).
Consequently, we fix the numbers $\kappa_\alpha$ as
\begin{eqnarray}
\kappa_\alpha = \expval{B_\alpha}\,,
\end{eqnarray}
where the expectation value has to be taken with respect to a thermal equilibrium state of the reservoir corresponding to $B_\alpha$.

With this convention, the total Hamiltonian can then be written as 
$\bar{\mathcal{H}} = \bar{\mathcal{H}}_{\rm S} + \bar{\mathcal{H}}_{\rm B} + \bar{\mathcal{H}}_{\rm SB}$.

Most simple, the reservoir part of the Hamiltonian remains invariant
\begin{eqnarray}
\label{eq:HBpol}
\bar{\mathcal{H}}_{\rm B}\equiv\sum_k \sum_{\sigma\in\{\rm L,R\}} \upvarepsilon_{k,\sigma}c_{k,\sigma}^\dagger c_{k,\sigma}
+\sum_q\omega_{q}\aq^\dagger\aq\,.
\end{eqnarray}

The system contribution to the Hamiltonian now experiences modified parameters
\begin{eqnarray}\label{EQ:hamsystem}
\bar{\mathcal{H}}_{\rm S} &=& \etL\dL^\dagger\dL+\etR\dR^\dagger\dR+\bar{U}\dL^\dagger\dL\dR^\dagger\dR\nn
&&+(\ttc e^{-2\ii\Phi} \dL \dR^\dagger+\ttc^* e^{+2\ii\Phi} \dR \dL^\dagger)\,,
\end{eqnarray}
with renormalized on-site energy levels~\cite{jovchev2013a} 
\begin{eqnarray}
\label{eq:RenormEnergy}
\bar{\upvarepsilon}_{\sigma}\equiv\upvarepsilon_\sigma-\sum_q\frac{\abs{h_{q,\sigma}}^2}{\omega_{q}}
\end{eqnarray}
and renormalized Coulomb repulsion~\cite{haertle2011a}
\begin{eqnarray}
\label{eq:RenormCBrep}
\bar{U}&\equiv& U-\sum_q\frac{h_{q,\rm L}^\ast h_{q,\rm R}+h_{q,\rm L}h_{q,\rm R}^\ast}{\omega_{q}}\,.
\end{eqnarray}
We note that in the strong-coupling limit, attractive Coulomb interactions ($\bar{U}<0$) are in principle possible~\cite{alexandrov2003a,koch2007a}.
Furthermore, we observe that also the internal tunneling amplitude is renormalized 
\begin{eqnarray}
\label{eq:ttc}
\ttc\equiv\tc\kappa\,,
\end{eqnarray}
where the complex-valued $\kappa$ is defined by  
$\kappa\equiv\expval{e^{-\mathcal{B}_{\rm L}}e^{\mathcal{B}_{\rm R}}}$.
Assuming that the phonon reservoir in the polaron-transformed frame is in thermal equilibrium, 
it (see Appendix~\ref{App:ShiftFactor}) explicitly evaluates to
\begin{eqnarray}\label{EQ:kappashift}
\kappa =  e^{-\sum_q\frac{\abs{h_{q,{\rm L}}-h_{q, {\rm R}}}^2}{\omega_{q}^2}[\frac{1}{2}+n(\omega_q)]}e^{+\ii\Phi}\,,
\end{eqnarray}
containing the Bose-distribution $n_{\rm B}(\omega)=[e^{\beta_{\rm ph}\omega}-1]^{-1}$ with the inverse phonon bath temperature $\beta_{\rm ph}$.
Here, the phase $\Phi$ is defined via
\begin{eqnarray}\label{EQ:phasefactor}
\ii\Phi\equiv\com{\mathcal{B}_{\rm L},\mathcal{B}_{\rm R}}/2=\sum_{q}\frac{h_{q, {\rm R}}^\ast h_{q, {\rm L}} -h_{q,{\rm R}}h_{q,{\rm L}}^\ast}{2\omega_{q}^2}\,.
\end{eqnarray}

Finally, the interaction Hamiltonian $\bar{\mathcal{H}}_{\rm SB}\equiv\bar{\mathcal{H}}_{\rm V}+\bar{\mathcal{H}}_{\rm T}$ is made of 
two parts.
The first describes electronic transitions between system and leads
\begin{eqnarray}
\label{eq:HV}
\bar{\mathcal{H}}_{\rm V}&\equiv&\sum_{k}(t_{k,\rm L} \dL e^{-\dR^\dagger \dR \ii\Phi} e^{-\mathcal{B}_{\rm L}} c_{k,\rm L}^\dagger+{\rm h.c.})\nn
&&+\sum_{k}(t_{k,\rm R} \dR e^{+\dL^\dagger \dL \ii\Phi} e^{-\mathcal{B}_{\rm R}} c_{k,\rm R}^\dagger+{\rm h.c.})\,,
\end{eqnarray}
which are now accompanied by multiple phonon emissions or absorptions.
The second part describes transitions between left and right dots
\begin{eqnarray}\label{eq:HT}
\bar{\mathcal{H}}_{\rm T}&\equiv& T_{\rm c} e^{-2\ii\Phi} \dL \dR^\dagger \left(e^{-\mathcal{B}_{\rm L}}e^{+\mathcal{B}_{\rm R}}-\kappa\right)\nn
&&+T_{\rm c} e^{+2\ii\Phi} \dR \dL^\dagger \left(e^{-\mathcal{B}_{\rm R}}e^{+\mathcal{B}_{\rm L}}-\kappa^*\right)\,,
\end{eqnarray}
which are also dressed by multiple phonon excitations, see Eq.~(\ref{eq:B_alpha}).

\begin{figure}[htb]
\begin{center}
\includegraphics[width=0.48\textwidth,clip=true]{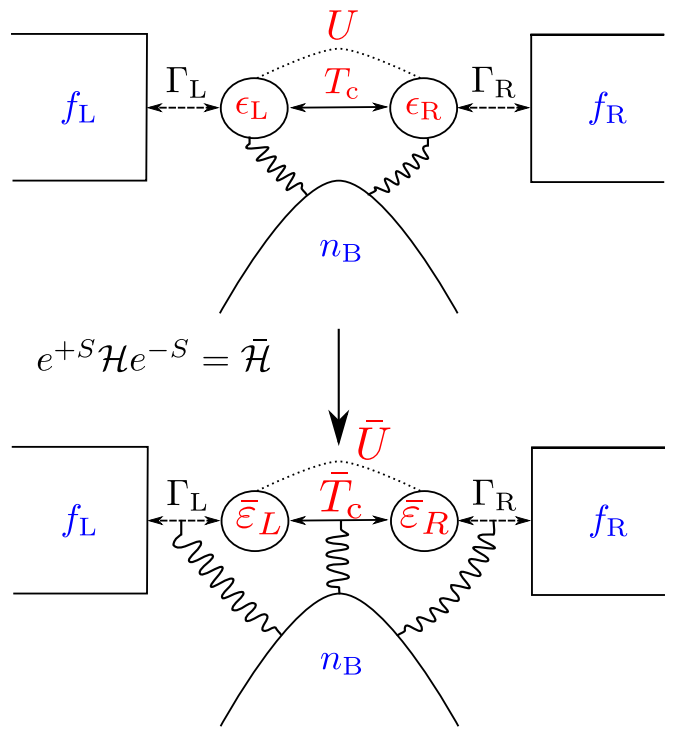}
\caption[]{\label{fig:PolTrafo} Sketch of the model before (top) and after
(below) the polaron transformation. The double quantum dot system in serial
configuration is coupled to electronic leads left and right each following
Fermi-Dirac statistics with Fermi functions $f_{\rm L}$ and $f_{\rm R}$,
respectively. If either temperatures or chemical potentials are chosen
differently, a non-equilibrium situation is created which enables the
exchange of matter and energy between those baths. The tunneling between
system and leads is described by the tunneling rates $\gL$ and $\gR$. The
quantum tunneling between left and right dot is modulated by the internal
tunneling rate $\tc$. Before the polaron transformation (with $S=\sum_\sigma d_\sigma^\dagger d_\sigma \mathcal{B}_\sigma$) 
the phonon bath with
Bose distribution $n_{\rm B}$ couples directly
to the occupation of the quantum dots left and right. Due to the polaron
transformation the coupling is shifted to the electronic jumps which now
occur with multiple phonon emission or absorption processes. Another feature of the model in
the polaron picture are the renormalized on-site energies and Coulomb repulsion
which now depend on the phonon coupling strength as well as the phonon mode frequency.}
\end{center}
\end{figure}

The effect of the polaron transformation is visualized in Fig.~\ref{fig:PolTrafo}.
The coupling to the phonon modes is no longer linear in the annihilation and creation operators anymore, 
as can be seen by expanding the exponentials $e^{\pm\mathcal{B}_\sigma}$. 
Comparing the system Hamiltonians before and after the polaron transformation, we see that 
apart from the renormalized on-site energies and Coulomb repulsion the electron-phonon interaction  
also renormalizes the internal tunneling term.
Consequently, the energy eigenbasis of  $\bar{\mathcal{H}}_{\rm S}$ is now influenced by the system-reservoir 
interaction strength in the original frame.

%%%%%%%%%%%%%%%%%%%%%%%%%%%%%%%%%%%%%%%%%%%%%%%%%%%%%%%%%%%%%%%%%%%%%%%%%%%%%%%
%%%%%%%%%%%%%%%%%%%%%%%%%%%%%%%%%%%%%%%%%%%%%%%%%%%%%%%%%%%%%%%%%%%%%%%%%%%%%%%

\subsection{Phonon treatment}\label{sec:differences}

We note that when the internal electronic tunneling amplitude was initially absent ($T_{\rm c}=0$), it would be straightforward
to keep the phonons as part of the system: Electron-phonon interactions would only arise from the
electronic jumps to and from the leads, such that diagonalizing the system Hamiltonian would be straightforward.
In this case, the resulting thermodynamics would be that of a two-terminal system exchanging matter and energy with the
two electronic leads.
For finite $T_{\rm c}$ however, keeping the phonons as part of the system would -- to obtain a thermodynamically
consistent master equation -- require to diagonalize an interacting infinite-dimensional Hamiltonian (such that
the polaron transformation would be of no use).
Therefore, we will proceed differently here and consider the phonons as part of the reservoir.
When we enforce the phonons in the polaron frame in a thermal equilibrium state 
$\propto e^{-\beta_{\rm ph} \bar{\mathcal H}_{\rm B}^{\rm ph}}$, this does in the
original frame actually correspond to a thermal phonon state that is conditioned on the
electronic occupation of the dots, see Appendix~\ref{App:polaron_inverse}.
A similar behaviour would be observed with phonons treated as part of the system, but
additionally strongly coupled to another thermal reservoir that imposes fast equilibration of
the phonons dependent on the electronic occupation~\cite{koch2006,muehlbacher2008a,simine2013a}.
We are aiming at a thermodynamically consistent description of this extreme limit, where the
phonons immediately equilibrate in an electron-dependent thermal state.

%%%%%%%%%%%%%%%%%%%%%%%%%%%%%%%%%%%%%%%%%%%%%%%%%%%%%%%%%%%%%%%%%%%%%%%%%%%%%%%
%%%%%%%%%%%%%%%%%%%%%%%%%%%%%%%%%%%%%%%%%%%%%%%%%%%%%%%%%%%%%%%%%%%%%%%%%%%%%%%
%%%%%%%%%%%%%%%%%%%%%%%%%%%%%%%%%%%%%%%%%%%%%%%%%%%%%%%%%%%%%%%%%%%%%%%%%%%%%%%
%%%%%%%%%%%%%%%%%%%%%%%%%%%%%%%%%%%%%%%%%%%%%%%%%%%%%%%%%%%%%%%%%%%%%%%%%%%%%%%

\section{Master equation in the strong electron-phonon coupling limit}\label{sec:MasterEquation}

%%%%%%%%%%%%%%%%%%%%%%%%%%%%%%%%%%%%%%%%%%%%%%%%%%%%%%%%%%%%%%%%%%%%%%%%%%%%%%%
%%%%%%%%%%%%%%%%%%%%%%%%%%%%%%%%%%%%%%%%%%%%%%%%%%%%%%%%%%%%%%%%%%%%%%%%%%%%%%%

\subsection{Pointer Basis}

We do now follow the standard derivation of a master equation~\cite{breuer2002,schaller2014}, starting from the general 
decomposition of the interaction Hamiltonian into system ($A_\alpha$) and bath ($B_\alpha$) operators (here in the Schr\"odinger picture)
\begin{eqnarray}
\label{eq:HSBdecomp}
\mathcal{H}_{\rm SB}=\sum_\alpha A_\alpha \otimes B_\alpha\,.
\end{eqnarray}
We note that such a tensor product decomposition is possible also for fermionic tunneling terms 
since one can map the fermionic operators to system and lead fermions via a Jordan-Wigner transform~\cite{schaller2009b}.
Ordering system and bath operators according to Eq.~(\ref{eq:HSBdecomp}), respectively, we obtain 6 coupling operators for system
\begin{eqnarray}
\label{eq:SystOp}
A_1&=&\dL=(A_2)^\dagger\,,\quad A_3=\dR=(A_4)^\dagger\,,\nn
A_5&=& e^{-2\ii\Phi} \dL \dR^\dagger=(A_6)^\dagger\,,
\end{eqnarray}
and reservoir
\begin{eqnarray}
\label{eq:BathOp}
B_1 &=& \sum_k t_{k,\rm L} c_{k,\rm L}^\dagger e^{-\mathcal{B}_{\rm L}}=B_2^\dagger\,,\nn
B_3 &=& \sum_k t_{k,\rm R} c_{k,\rm R}^\dagger e^{-\mathcal{B}_{\rm R}}=B_4^\dagger\,,\nn
B_5 &=& e^{-\mathcal{B}_{\rm L}}e^{+\mathcal{B}_{\rm R}}-\kappa=B_6^\dagger\,.
\end{eqnarray}
The expectation value of two bath operators defines the bath correlation function
\begin{eqnarray}
\label{eq:BCF}
\mathcal{C}_{\alpha\beta}(\tau)\equiv\expval{{\f B}_\alpha(\tau){\f B}_\beta(0)}\,,
\end{eqnarray}
where bold symbols denote the interaction picture ${\f B}_\alpha(\tau) = e^{+\ii \bar\HB \tau} B_\alpha e^{-\ii \bar\HB \tau}$
and where the reservoir $\RB = \RB^{\rm L} \otimes \RB^{\rm R} \otimes \RB^{\rm ph}$ is a tensor product of 
thermalized states of left and right electronic leads and the phonon reservoir, respectively.
This simple tensor-product approximation in the polaron-transformed frame does not hold in the original frame, 
where one obtains a displaced thermal phonon state depending on
the electronic occupations, which is explicitly shown in appendix~\ref{App:polaron_inverse}.

When the electronic reservoirs are weakly coupled and sufficiently Markovian (formalized by the condition $\beta_\alpha \Gamma_\alpha \ll 1$), 
perturbation theory in the electron-lead tunneling amplitudes $t_{k\sigma}$ and in $\tc (\kappa-1)$, i.e., 
either in the asymmetry of the electron-phonon coupling or in $\tc$ itself 
(for a continuum of phonon modes we just require a perturbative treatment in the $t_{k\sigma}$),
leads to a Lindblad master equation, which appears particularly simple in the system energy eigenbasis.
We label the eigenvectors of $\bar\HS$ as
$\ket{a}\in\left\{\ket{0},\ket{-},\ket{+},\ket{2}\right\}$, with system eigenenergies
\begin{eqnarray}
\label{eq:systEigEnergies}
\eO&\equiv&0\,,\\
\eM&\equiv&\frac{1}{2} \left(\etL+\etR-\sqrt{(\etL-\etR)^2+4\abs{\ttc}^2}\right)\,,\nn
\eP&\equiv&\frac{1}{2} \left(\etL+\etR+\sqrt{(\etL-\etR)^2+4\abs{\ttc}^2}\right)\,,\nn
\ePM&\equiv&\etL+\etR+\bar U\,.
\end{eqnarray}
When the system spectrum is non-degenerate (more precisely, when the splitting $\eP-\eM$ in $\bar\HS$ is much larger than the electronic 
tunneling amplitudes in the Hamiltonian), this will lead to a rate equation for the populations in the
system energy eigenbasis
\begin{eqnarray}
\dot\rho_{aa} = +\sum_b \gamma_{ab,ab} \rho_{bb} - \left[\sum_b \gamma_{ba,ba}\right] \rho_{aa}\,,
\end{eqnarray}
where the positive rates~\cite{schaller2014}
\begin{eqnarray}
\gamma_{ab,ab} = \sum_{\alpha\beta} \gamma_{\alpha\beta}(\epsilon_b-\epsilon_a) \bra{a} A_\beta\ket{b} \bra{a} A_\alpha^\dagger \ket{b}^*\,.
\end{eqnarray}
are given by matrix elements of the system coupling operators in the energy eigenbasis and the Fourier transform of the reservoir correlation functions
\begin{eqnarray}
\label{eq:rate}
\gamma_{\alpha\beta}(\omega) = \int dt \, e^{+\ii \omega t} \mathcal{C}_{\alpha\beta}(t)\,.
\end{eqnarray}
The coherences evolve independently from the populations.
In particular, since in our model only coherences between states with the same charge may exist, we have
\begin{eqnarray}\label{EQ:coherenceevolution}
\dot{\rho}_{-+} &=& -\ii\left(E_--E_++\sigma_{--}-\sigma_{++}\right) \rho_{-+}\nn
&&+\Big[\gamma_{--,++}-\frac{\gamma_{0-,0-}+\gamma_{0+,0+}+\gamma_{2-,2-}+\gamma_{2+,2+}}{2}\nn
&&-\frac{\gamma_{-+,-+}+\gamma_{+-,+-}}{2}\Big] \rho_{-+}\,,
\end{eqnarray}
%where $\sigma_{ii}\in\mathbb{R}$ describes a level-renormalization (Lamb-shift).
where $\sigma_{--},\sigma_{++}\in\mathbb{R}$ describe a level-renormalization (Lamb-shift).
%
%We note that we can additively decompose the rates from $j$ to $i$ involving electronic exchange
We note that the rates $\gamma_{ab,ab}$ in Eq.~(\ref{EQ:coherenceevolution}) which describe electronic tunneling with phononic excitation between system and leads can be decomposed into interaction with left (L) and right (R) bath
\begin{eqnarray}\label{EQ:defrateel}
\gamma_{ab,ab} \equiv \Gamma_{\rm L}^{ab} + \Gamma_{\rm R}^{ab}\,,
\end{eqnarray}
whereas the internal electronic transitions $\gamma_{-+,-+}$ and $\gamma_{+-,+-}$ describing the relaxation from $+$ to $-$ or the excitation from $-$ to $+$, respectively, only involve phonons
\begin{eqnarray}\label{EQ:defrateph}
\gamma_{-+,-+} \equiv \Gamma_{\rm ph}^{-+}\,,\qquad
\gamma_{+-,+-} \equiv \Gamma_{\rm ph}^{+-}\,.
\end{eqnarray}
Furthermore, the matrix elements in the rates describing backward and forward processes triggered by the same reservoir are identical,
such that local detailed balance is only induced by a corresponding Kubo-Martin-Schwinger (KMS)-type condition of the correlation functions.
We discuss these for our system in Sec.~\ref{sec:LeadPhononBCF} and Sec.~\ref{sec:dotphononbcf}.

As a distinctive feature in comparison to a single quantum dot~\cite{schaller2013a} or to models
without internal phonon-independent electronic tunneling, one now obtains phonon-modified internal transitions, 
and the corresponding rates between energy eigenstates $\ket{-}$ and $\ket{+}$ 
can be written as a quadratic form
\begin{eqnarray}\label{eq:Abbreviationsmp}
\Gamma_{\rm ph}^{-+}&=&
\left(A_5^{+-},(A_5^{-+})^\ast\right)
\underline{\gamma(\eP-\eM)}
\begin{pmatrix}
 (A_5^{+-})^\ast \\
 A_5^{-+}
\end{pmatrix}\,,\nn
\Gamma_{\rm ph}^{+-}&=&
\left(A_5^{-+},(A_5^{+-})^\ast\right)
\underline{\gamma(\eM-\eP)}
\begin{pmatrix}
 (A_5^{-+})^\ast \\
 A_5^{+-}
\end{pmatrix}
\end{eqnarray}
with the matrix $\underline{\gamma(\omega)}$ being given by
\begin{eqnarray}\label{EQ:phonon_matrix}
\underline{\gamma(\omega)}=
\begin{pmatrix}
\gamma_{56}(\omega) & \gamma_{55}(\omega) \\
\gamma_{66}(\omega) & \gamma_{65}(\omega)
\end{pmatrix}\,.
\end{eqnarray}
It can be shown that this matrix is hermitian and positive definite, such that we 
obtain true rates $\Gamma_{\rm ph}^{-+}\ge 0$ and $\Gamma_{\rm ph}^{+-}\ge 0$.
Furthermore, we note that since the correlation functions contained in the matrix~(\ref{EQ:phonon_matrix})
obey KMS relations of the form $\gamma_{\alpha\beta}(-\omega)=\gamma_{\beta\alpha}(+\omega) e^{-\beta_{\rm ph} \omega}$ with
inverse phonon reservoir temperature $\beta_{\rm ph}$ (compare Sec.~\ref{sec:dotphononbcf}), 
this implies for the ratio of rates $\frac{\Gamma_{\rm ph}^{+-}}{\Gamma_{\rm ph}^{-+}} = e^{-\beta_{\rm ph} (\eP-\eM)}$.

%%%%%%%%%%%%%%%%%%%%%%%%%%%%%%%%%%%%%%%%%%%%%%%%%%%%%%%%%%%%%%%%%%%%%%%%%%%%%%%
%%%%%%%%%%%%%%%%%%%%%%%%%%%%%%%%%%%%%%%%%%%%%%%%%%%%%%%%%%%%%%%%%%%%%%%%%%%%%%%

\subsection{Lead-Phonon Correlation Function}\label{sec:LeadPhononBCF}

From Eq.~(\ref{eq:BCF}) it follows that the four non-vanishing contributions associated with electronic jumps into or out of the 
system can be written in a product form of electronic and phononic contributions~\cite{schaller2013a}
\begin{eqnarray}
\mathcal{C}_{\alpha\beta}(\tau)=\mathcal{C}_{\alpha\beta}^{\rm el}(\tau)\mathcal{C}_{\alpha\beta}^{\rm ph}(\tau)\,,
\end{eqnarray}
with the electronic parts being given by
\begin{eqnarray}
\mathcal{C}_{12}^{\rm el}(\tau) &=& \sum_k \abs{\tkL}^2 f_{\rm L}(\ekL) e^{+\ii\ekL\tau}\,,\nn
\mathcal{C}_{21}^{\rm el}(\tau) &=& \sum_k \abs{\tkL}^2 [1-f_{\rm L}(\ekL)] e^{-\ii\ekL\tau}\,,\nn
\mathcal{C}_{34}^{\rm el}(\tau) &=& \sum_k \abs{\tkR}^2 f_{\rm R}(\ekR) e^{+\ii\ekR\tau}\,,\nn
\mathcal{C}_{43}^{\rm el}(\tau) &=& \sum_k \abs{\tkR}^2 [1-f_{\rm R}(\ekR)] e^{-\ii\ekR\tau}\,.
\end{eqnarray}
Here, we have introduced the Fermi function $f_\sigma(\omega)\equiv [e^{\beta_\sigma(\omega-\mu_\sigma)}+1]^{-1}$ of 
lead $\sigma$ with inverse temperature $\beta_\sigma$ and chemical potential $\mu_\sigma$.
The tunneling amplitudes $t_{k\sigma}$ lead to effective tunneling rates $\Gamma_\sigma(\omega) \equiv 2\pi\sum_k \abs{t_{k\sigma}}^2\delta(\omega-\upvarepsilon_{k\sigma})$, 
which can be used to convert the above summations into integrals.
Later-on, we will parametrize them with a Lorentzian distribution~\cite{zedler2009a}
\begin{eqnarray}
\Gamma_\sigma(\omega)\equiv\frac{\Gamma_\sigma \delta_\sigma^2}{\omega^2+\delta_\sigma^2}\,,
\end{eqnarray}
yielding a representation in terms of hypergeometric functions for $C_{\alpha\beta}^{\rm el}(\tau)$,
which we omit here for brevity. 
For completeness we note that the separate Fourier transforms of the electronic 
parts $\gamma_{\alpha\beta}^{\rm el}(\omega) = \int \mathcal{C}_{\alpha\beta}^{\rm el}(\tau) e^{+\ii\omega\tau} d\tau$
\begin{eqnarray}
\gamma_{12}^{\rm el}(\omega) &=& \Gamma_{\rm L}(-\omega) f_{\rm L}(-\omega)\,,\nn
\gamma_{21}^{\rm el}(\omega) &=& \Gamma_{\rm L}(+\omega)[1-f_{\rm L}(+\omega)]\,,\nn
\gamma_{34}^{\rm el}(\omega) &=& \Gamma_{\rm R}(-\omega) f_{\rm R}(-\omega)\,,\nn
\gamma_{43}^{\rm el}(\omega) &=& \Gamma_{\rm R}(+\omega)[1-f_{\rm R}(+\omega)]\,,
\end{eqnarray}
obey -- since $f_\sigma(\omega) = e^{-\beta_\sigma(\omega-\mu_\sigma)}[1-f_\sigma(\omega)]$ -- the KMS-type relations
\begin{eqnarray}\label{EQ:kmselectronic}
\gamma_{12}^{\rm el}(-\omega) &=& e^{-\beta_{\rm L}(\omega-\mu_{\rm L})} \gamma_{21}^{\rm el}(+\omega)\,,\nn
\gamma_{34}^{\rm el}(-\omega) &=& e^{-\beta_{\rm R}(\omega-\mu_{\rm R})} \gamma_{43}^{\rm el}(+\omega)\,.
\end{eqnarray}

The phonon contribution to the correlation function depends only on the terminal across which
the electron jumps but not on the jump direction, i.e., we have
$\mathcal{C}_{12}^{\rm ph}(\tau) = \mathcal{C}_{21}^{\rm ph}(\tau) \equiv \mathcal{C}_{\rm L}^{\rm ph}(\tau)$ and
$\mathcal{C}_{34}^{\rm ph}(\tau) = \mathcal{C}_{43}^{\rm ph}(\tau) \equiv \mathcal{C}_{\rm R}^{\rm ph}(\tau)$.
Using the Baker-Campbell-Hausdorff (BCH) formula, 
the phonon contribution explicitly computes to (see Appendix~\ref{App:PhononBCF})
\begin{eqnarray}
\label{eq:BCFphonon}
\mathcal{C}_\sigma^{\rm ph}(\tau)=e^{-K_\sigma(0)+K_\sigma(\tau)}\,,
\end{eqnarray}
with the abbreviation in the exponent
\begin{eqnarray}
\label{eq:PhononBCFexponent1}
K_\sigma(\tau)&=&\sum_q\frac{\abs{\hqa}^2}{\omega_q^2}\times\\
&&\times\left\lbrace n_{\rm B}(\omega_q)e^{+\ii\omega_q\tau}+[n_{\rm B}(\omega_q)+1]e^{-\ii\omega_q\tau}\right\rbrace\,.\nonumber
\end{eqnarray}
It is easy to show that $K_\sigma(\tau)=K_\sigma(-\tau-\ii\beta_{\rm ph})$ holds, which transfers to the KMS condition for the phonon
contribution to the correlation function 
\begin{eqnarray}\label{EQ:kmsphononic}
\mathcal{C}^{\rm ph}_{\sigma}(\tau)=\mathcal{C}^{\rm ph}_{\sigma}(-\tau-\ii\beta_{\rm ph})\,.
\end{eqnarray}

The nature of the phonon contributions can now be quite distinct depending on whether one has a discrete (e.g.\  just a single mode) 
or continuous spectrum of phonon frequencies.
In the continuum case, we can convert the sum in the exponent into an integral. 
Then, the phonon absorption emission amplitudes enter the corresponding 
rate as $\mathcal{J}_\sigma(\omega)\equiv\sum_q\abs{h_{q\sigma}}^2\delta(\omega-\omega_q)$, where 
$\mathcal{J}_{\rm L}(\omega)$ and $\mathcal{J}_{\rm R}(\omega)$ will be parametrized by
a continuous function.
For example, using the super-ohmic parameterization with exponential infrared cut-off at $\wc^\sigma$ (we choose a super-ohmic representation
to enable a Markovian description of the internal jumps in Sec.~\ref{sec:dotphononbcf}) and coupling strength $J_\sigma$, i.e.,
\begin{eqnarray}
\mathcal{J}_\sigma(\omega)\equiv J_\sigma \omega^3 e^{-\frac{\omega}{\wc^\sigma}}\,,
\end{eqnarray}
we obtain for the integrals in the exponent
\begin{eqnarray}\label{EQ:expcorrfunc}
K_\sigma(\tau) &=& \int\limits_0^\infty \frac{{\mathcal J}_\sigma(\omega)}{\omega^2} \left[n_{\rm B}(\omega) e^{+\ii\omega\tau} + [1+n_{\rm B}(\omega)] e^{-\ii\omega\tau}\right] d\omega\nn
&=& \frac{2 J_\sigma}{\beta^2} \Re\left\{\Psi'\left(\frac{1+\ii\tau\wc^\sigma}{\bPH\wc^\sigma}\right)\right\} 
- \frac{J_\sigma (\wc^\sigma)^2}{\left(1-\ii\tau\wc^\sigma\right)^2}\,,
\end{eqnarray}
where $\Psi'(x)$ denotes the derivative of the PolyGamma function $\Psi(x) = \Gamma'(x)/\Gamma(x)$.
With the same super-ohmic spectral density, the renormalized on-site energies and Coulomb shift read explicitly
\begin{eqnarray}
\bar\upvarepsilon_\sigma &=& \upvarepsilon_\sigma - 2 J_\sigma (\wc^\sigma)^3\,,\nn
\bar U &=& U + \sum_q \frac{\abs{\hqL-\hqR}^2-\abs{\hqL}^2-\abs{\hqR}^2}{\omega_q}\nn
&=& U +  2 J_0 (\wc^0)^3 - 2 J_{\rm L} (\wc^{\rm L})^3 - 2 J_{\rm R} (\wc^{\rm R})^3\,.
\end{eqnarray}
We note here that since $K_\sigma(\tau)$ in Eq.~(\ref{EQ:expcorrfunc}) decays to zero for large $\tau$, the phonon correlation 
function $C_\sigma^{\rm ph}(\tau)$ may remain finite for large $\tau$.
Thanks to the influence of the electronic contributions the total correlation function will still decay, 
such that a Markovian approach is applicable.
In this case we technically define separate Fourier transforms of the phonon contributions by 
\begin{eqnarray}
\gamma_\sigma^{\rm ph}(\omega) &=& \int \left[\mathcal{C}_\sigma^{\rm ph}(\tau) - \mathcal{C}_\sigma^{\rm ph}(\infty)\right] e^{+\ii\omega\tau} d\tau\nn
&&+ 2\pi \mathcal{C}_\sigma^{\rm ph}(\infty) \delta(\omega)\,.
\end{eqnarray}

Since the dressed correlation functions are given by products of electronic and phononic contributions in the time domain, 
the separate KMS relations~(\ref{EQ:kmselectronic}) and~(\ref{EQ:kmsphononic}) do not directly transfer in non-equilibrium setups.
However, we can use our previous result (see appendix of Ref.~\cite{schaller2013a}) that these correlation functions can be
written conditioned upon the net number $\f{n}=(n_1,\ldots,n_Q)$ of emitted phonons into the different reservoir
modes ($n_q<0$ implies absorption from the phonon reservoir).
Formally, one has 
$\gamma_{\alpha\beta}(\omega)=\sum_{\f{n}} \gamma_{\alpha\beta,\f{n}}(\omega)$,
where the separate contributions are given by ($\f{\Omega}=(\omega_1,\ldots,\omega_Q)$)
\begin{eqnarray}\label{eq:RateNdep}
\gamma_{\alpha\beta,\f{n}}(\omega)&=&\gamma_{\alpha\beta}^{\rm el}(\omega-\f{n} \cdot\f{\Omega}) \prod_q e^{-\frac{\abs{h_q}^2}{\omega_q^2}(1+2 n_{\rm B}^q)} \times\nn
&&\times \left(\frac{1+\nB^q}{\nB^q}\right)^{n_q/2}\times\nn
&&\times \mathcal{J}_{n_q}\left(2\frac{\abs{h_q}^2}{\omega_q^2}\sqrt{\nB^q(1+\nB^q)}\right)\,,
\end{eqnarray}
with $\mathcal{J}_n(x)\equiv\sum_{k=0}^\infty \lbrace(-1)^k/k!\Gamma[k+n+1]\rbrace (x/2)^{2k+n}$ being the modified
Bessel function of the first kind and $\Gamma[x]\equiv\int_0^\infty t^{x-1}e^{-t}{\rm d}t$ being the Gamma-function.
We note that when the electronic Fourier transforms are flat $\gamma_{\alpha\beta}^{\rm el}(\omega-\f{n} \cdot\f{\Omega})\to\bar\gamma_{\alpha\beta}^{\rm el}$, 
the normalization of the phonon contribution implies that the Fourier transform of the combined
correlation function is also flat $\sum_{\f{n}} \gamma_{\alpha\beta,\f{n}}(\omega)\to\bar\gamma_{\alpha\beta}^{\rm el}$.
This implies that in the electronic wide-band ($\delta_\sigma\to\infty$) plus the infinite bias ($f_{\rm L}(\omega)\to 1$ and $f_{\rm R}(\omega)\to 0$) limits
the phonons will have no effect on the dot-lead correlation functions.

Importantly, we note that even for different temperatures, these obey the KMS-type relation
\begin{eqnarray}
\gamma_{12,+\f{n_{\rm L}}}(-\omega) &=& e^{-\bL(\omega-\mL+\f{n_{\rm L}}\cdot \f{\Omega})}e^{+\beta_{\rm ph} \f{n_{\rm L}} \cdot \f{\Omega}}\times\nn
&&\times \gamma_{21,-\f{n_{\rm L}}}(+\omega)\,,\nn
\gamma_{34,+\f{n_{\rm R}}}(-\omega) &=& e^{-\bR(\omega-\mR+\f{n_{\rm R}}\cdot \f{\Omega})}e^{+\beta_{\rm ph} \f{n_{\rm R}} \cdot \f{\Omega}}\times\nn
&&\times \gamma_{43,-\f{n_{\rm R}}}(+\omega)\,.
\end{eqnarray}
We see that the conventional KMS relation is reproduced when phonon and electronic temperatures are equal.

%%%%%%%%%%%%%%%%%%%%%%%%%%%%%%%%%%%%%%%%%%%%%%%%%%%%%%%%%%%%%%%%%%%%%%%%%%%%%%%
%%%%%%%%%%%%%%%%%%%%%%%%%%%%%%%%%%%%%%%%%%%%%%%%%%%%%%%%%%%%%%%%%%%%%%%%%%%%%%%

\subsection{Interdot-Phonon Correlation Function}\label{sec:dotphononbcf}

To evaluate the transitions 
between the states $\ket{-}\leftrightarrow\ket{+}$, we have to evaluate the correlation functions
\begin{eqnarray}
\mathcal{C}_{55}(\tau) 
&=&\expval{e^{-\boldsymbol{\mathcal{B}}_{\rm L}(\tau)}e^{+\boldsymbol{\mathcal{B}}_{\rm R}(\tau)}e^{-\mathcal{B}_{\rm L}}e^{+\mathcal{B}_{\rm R}}}-\kappa^{2}\,,\nn
\mathcal{C}_{66}(\tau) 
&=&\expval{e^{-\boldsymbol{\mathcal{B}}_{\rm R}(\tau)}e^{+\boldsymbol{\mathcal{B}}_{\rm L}(\tau)}e^{-\mathcal{B}_{\rm R}}e^{+\mathcal{B}_{\rm L}}}-(\kappa^\ast)^{2}\,,\nn
\mathcal{C}_{56}(\tau) 
&=&\expval{e^{-\boldsymbol{\mathcal{B}}_{\rm L}(\tau)}e^{+\boldsymbol{\mathcal{B}}_{\rm R}(\tau)}e^{-\mathcal{B}_{\rm R}}e^{+\mathcal{B}_{\rm L}}}-\abs{\kappa}^{2}\,,\nn
\mathcal{C}_{65}(\tau) 
&=&\expval{e^{-\boldsymbol{\mathcal{B}}_{\rm R}(\tau)}e^{+\boldsymbol{\mathcal{B}}_{\rm L}(\tau)}e^{-\mathcal{B}_{\rm L}}e^{+\mathcal{B}_{\rm R}}}-\abs{\kappa}^{2}\,,
\end{eqnarray}
where we have used that $\kappa = \expval{e^{-\mathcal{B}_{\rm L}}e^{+\mathcal{B}_{\rm R}}} = \expval{e^{-\boldsymbol{\mathcal{B}}_{\rm L}(\tau)}e^{+\boldsymbol{\mathcal{B}}_{\rm R}(\tau)}}$
is inert with respect to transformations into the interaction picture.
For the first bath correlation functions we obtain (see Appendix~\ref{App:InterdotBCF})
\begin{eqnarray}
\label{eq:C55}
\mathcal{C}_{55}(\tau)=\kappa^2\left[e^{-K(\tau)}-1\right]\,,
\end{eqnarray}
where -- in analogy to Eq.~(\ref{eq:PhononBCFexponent1}) -- we have
\begin{eqnarray}\label{eq:PhononBCFexponent2}
K(\tau) &=&\sum_q\frac{\abs{\hqL-\hqR}^2}{\omega_q^2}\times\\
&&\times\left\lbrace n_{\rm B}(\omega_q)e^{+\ii\omega_q\tau}+[n_{\rm B}(\omega_q)+1]e^{-\ii\omega_q\tau}\right\rbrace\,.\nonumber
\end{eqnarray}
We note that for large times the correlation function vanishes for a continuum of phonon modes, facilitating a Markovian description.
Two further correlation functions can be similarly evaluated
\begin{eqnarray}
\label{eq:C66}
\mathcal{C}_{66}(\tau)=(\kappa^\ast)^2\left[e^{-K(\tau)}-1\right]=\mathcal{C}_{55}^\ast(-\tau)\,,
\end{eqnarray}
where the latter equality can be easily seen by direct comparison. 
For the third correlation function we find
\begin{eqnarray}
\label{eq:C56}
&&\mathcal{C}_{56}(\tau)=\abs{\kappa}^2\left[e^{K(\tau)}-1\right]\,.
\end{eqnarray}
It can be easily seen that $\mathcal{C}_{56}(t)\hat{=}\mathcal{C}_{65}(t)$.
Furthermore, we note that
\begin{eqnarray}
\kappa^2&=&e^{-K(0)}e^{+2\ii\Phi}\,,\quad (\kappa^\ast)^2=e^{-K(0)}e^{-2\ii\Phi}\,,\nn
\abs{\kappa}^2&=&e^{-K(0)}\,.
\end{eqnarray}
From $K(-\tau)=K^\ast(+\tau)$ we conclude that the Fourier transform matrix of these correlation functions~(\ref{EQ:phonon_matrix}) is hermitian.
It can be expressed by the two real-valued functions
\begin{eqnarray}
\gamma_\pm(\omega)=\int\left(e^{\pm K(\tau)}-1\right)e^{+\ii \omega \tau}d\tau
\end{eqnarray}
and will be positive definite at frequency $\omega$ when $\gamma_-(\omega)<\gamma_+(\omega)$ or, 
equivalently, when $\gamma^2_+(\omega)-\gamma^2_-(\omega)=[\gamma_+(\omega)-\gamma_-(\omega)][\gamma_+(\omega)+\gamma_-(\omega)]>0$.
The interdot phonon correlation functions obey KMS relations of the type (for $\alpha,\beta\in\{5,6\}$)
\begin{eqnarray}
C_{\alpha\beta}(\tau) = C_{\beta\alpha}(-\tau-\ii\beta_{\rm ph})\,,
\end{eqnarray}
which follow from the definition of $K(\tau)$.
For their Fourier transforms this implies $\gamma_{\alpha\beta}(-\omega) = \gamma_{\beta\alpha}(+\omega) e^{-\beta\omega}$.

Finally, we note that this approach is valid for coupling to a continuum of phonon modes.
A finite number of phonon modes would in general 
not lead to a decay of the inter-dot correlation functions $\mathcal{C}_{55}(\tau)$, $\mathcal{C}_{56}(\tau)$, $\mathcal{C}_{65}(\tau)$, and $\mathcal{C}_{66}(\tau)$, 
thus prohibiting a Markovian description.
Furthermore, the electronic tunneling Hamiltonian $\bar{\mathcal{H}}_{\rm V}$ and the inter-dot tunneling Hamiltonian $\bar{\mathcal{H}}_{\rm T}$ must be small
in the polaron frame.
The first condition is consistent with a perturbative treatment of electron-lead tunneling amplitudes, whereas the second
condition can be fulfilled by choosing either 
nearly symmetric electron-phonon couplings left and right, i.e. $\hqL\approx h_q \approx \hqR$
or by treating $\tc$ also perturbatively.
If the electron-phonon coupling is exactly symmetric, also finite phonon modes can be treated with the approach.

%%%%%%%%%%%%%%%%%%%%%%%%%%%%%%%%%%%%%%%%%%%%%%%%%%%%%%%%%%%%%%%%%%%%%%%%%%%%%%%
%%%%%%%%%%%%%%%%%%%%%%%%%%%%%%%%%%%%%%%%%%%%%%%%%%%%%%%%%%%%%%%%%%%%%%%%%%%%%%%

\subsection{Numerical phonon correlation function}\label{sec:phonon_numeric}

In case of a continuous phonon spectrum, the Fourier transforms of the phonon 
correlation functions associated with external -- compare Eq.~(\ref{eq:BCFphonon}) -- and
internal -- compare Eqns.~(\ref{eq:C55}),~(\ref{eq:C66}), and~(\ref{eq:C56}) -- electronic jumps
cannot be obtained analytically in closed form.
This complicates the calculation of the full transition rates whenever one is also interested in the heat
exchanged with the phonon reservoir, as this requires evaluation of a convolution integral, where the
phonon contribution to the integrand is itself a numerical Fourier integral.
Here, we therefore aim to represent the Fourier-transform of the phonon contribution in a semi-exact fashion, 
respecting the thermodynamic KMS relations.
For this, we note that the Gaussian
\begin{eqnarray}\label{eq:gaussfitf}
\gamma_{\rm ph}^{\rm fit}(\omega) = a e^{-\frac{(\omega-\beta_{\rm ph} b/4)^2}{b}}
\end{eqnarray}
obeys for all fit parameters $a$ and $b$ and frequencies $\omega$ the KMS relation 
$\frac{\gamma_{\rm ph}^{\rm fit}(+\omega)}{\gamma_{\rm ph}^{\rm fit}(-\omega)} = e^{\beta_{\rm ph}\omega}$, 
where $\beta_{\rm ph}$ denotes the inverse phonon temperature.
Naturally, by fitting the phonon correlation functions e.g.\  with multiple such Gaussian functions one
would obtain a thermodynamic correct representation of the phonon correlation function.
Here however, we are rather interested in thermodynamic principles and just use a single Gaussian function, where we fix the
fit parameters by crudely matching
$C_{\rm ph}^{\rm fit}(0)$ and $\int C_{\rm ph}^{\rm fit}(\tau) d\tau$ with the true values of the correlation function.
We note that both $C_{\rm ph}(0)$ and $\int C_{\rm ph}(\tau) d\tau$ are always real-valued, such that the Fourier transform of
the Gaussian approximation does not only obey the KMS condition but is also always positive.

%%%%%%%%%%%%%%%%%%%%%%%%%%%%%%%%%%%%%%%%%%%%%%%%%%%%%%%%%%%%%%%%%%%%%%%%%%%%%%%
%%%%%%%%%%%%%%%%%%%%%%%%%%%%%%%%%%%%%%%%%%%%%%%%%%%%%%%%%%%%%%%%%%%%%%%%%%%%%%%

\section{Symmetries in the Full Counting Statistics}\label{sec:FCSsingleMode}

To deduce the counting statistics not only of electrons but also of the phonons, it would be
necessary to identify the phonons emitted or absorbed with every electronic jump.
However, here we are rather interested in the energy that by such processes is emitted into or
absorbed from the phonon reservoir.
For internal electronic transitions, the energy exchange follows directly from the change in the
system state.
In contrast, for transitions involving an electronic jump across the left or right terminal, 
one has to identify the separate phononic contributions to correctly partition the electronic
and phononic contributions to the exchanged energy.

To identify a minimal set of transitions that has to be monitored for energy and particle exchange, 
we first consider the entropy production $\dot{S}_\ii$ in the system, which at steady state must
be balanced by the entropy flow $\dot{S}_{\rm e}$ from the electronic and phononic terminals~\cite{esposito2010a}
\begin{eqnarray}
\dot{S}_\ii &=& -\dot{S}_{\rm e} = -\sum_\nu \beta_\nu \dot{Q}_\nu\nn
&=& -\bL(I_E^{\rm L}-\mL I_M^{\rm L})-\bR(I_E^{\rm R}-\mR I_M^{\rm R})\nn
&&-\beta_{\rm ph} I_E^{\rm ph}\,,
\end{eqnarray}
where $I_E^\nu$, $I_M^\nu$, and $\dot{Q}_\nu$ denote the energy, matter, and heat currents from terminal $\nu$ into the system, respectively.
Using the conservation laws for energy and matter
\begin{eqnarray}
I_{\rm E}^{\rm L}+I_{\rm E}^{\rm R}+I_{\rm E}^{\rm ph}&=&0\,,\qquad
I_{\rm M}^{\rm L}+I_{\rm M}^{\rm R}=0\,,
\end{eqnarray}
we can eliminate two currents.

We choose to monitor the number of electrons entering the system from the left lead $I_{\rm M}^{(\rm L)}$, 
the energy that is transferred from the left lead into the system $I_{\rm E}^{(\rm L)}$, and the energy
that is transferred from the phonon reservoir into the system $I_{\rm E}^{(\rm ph)}$.
In terms of these quantities, the entropy production becomes
\begin{eqnarray}\label{EQ:entprodfull}
\dot{S}_\ii &=& (\bR-\bL) I_{\rm E}^{\rm L} + (\bL \mL - \bR \mR) I_{\rm M}^{\rm L}\nn
&& + (\bR - \beta_{\rm ph}) I_{\rm E}^{\rm ph}\,,
\end{eqnarray}
which is decomposable into affinities and fluxes.
When we further assume that the electronic temperatures of both leads are the same $\bL=\bR=\beta_{\rm el}$, 
the entropy production can even be expressed with only two affinities and two fluxes
\begin{eqnarray}\label{EQ:entprodred}
\dot{S}_\ii &=& \beta_{\rm el} (\mL-\mR) I_{\rm M}^{\rm L} + (\beta_{\rm el} - \beta_{\rm ph}) I_{\rm E}^{\rm ph}\,.
\end{eqnarray}
Formally, the statistics of energy and matter transfers can be extracted 
by complementing the off-diagonal entries in the Liouvillian that
describe the individual jump processes with counting fields.
For the electronic hopping this is fairly standard and straightforward to do.
It becomes a bit more involved however when one is interested in the statistics
of energy exchanges:
For the internal jumps -- see Eq.~(\ref{EQ:defrateph}) -- the energy counting field $\phi$ is multiplied by the complete energy that is exchanged with the phonon reservoir
\begin{eqnarray}
\Gamma_{\rm ph}^{-+} &\to& \Gamma_{\rm ph}^{-+} e^{-\ii\phi(\eP-\eM)}\,,\nn
\Gamma_{\rm ph}^{+-} &\to& \Gamma_{\rm ph}^{+-} e^{+\ii\phi(\eP-\eM)}\,.
\end{eqnarray}
For the electronic jumps between system and both leads we however have to partition the emitted or absorbed energy
into contributions from the electronic and phononic reservoirs, which first requires to decompose the 
transitions into different phonon contributions.
Assuming for example a discrete phonon spectrum we have
\begin{eqnarray}
\Gamma_\sigma^{ab} = \sum_{\f{n}} \Gamma_\sigma^{ab,\f{n}}\,,
\end{eqnarray}
where $\Gamma_\sigma^{ab,\f{n}}$ describes a transition from energy eigenstate $j$ to $i$ together with the emission
of $\f{n}$ phonons into the different phonon reservoir modes and an electronic jump to or from lead $\sigma\in\{\rm L,R\}$ -- see Eq.~(\ref{EQ:defrateel}).
For a continuous phonon spectrum (which we will not discuss explicitly) we could use the convolution theorem to arrive at a similar 
decomposition $\Gamma_\sigma^{ab} = \int \Gamma_\sigma^{ab}(\omega) d\omega$, 
where $\Gamma_\sigma^{ab}(\omega)$ describes a transition from energy eigenstate $b$ to $a$ together with the emission
of energy $\omega$ into the phonon reservoir and an electronic jump to or from lead $\sigma\in\{\rm L,R\}$.
This then implies the counting field replacements for the off-diagonal matrix elements in the Liouvillian
\begin{eqnarray}
\gL^{ab,\f{n}} &\to& \gL^{ab,\f{n}} e^{+\ii\chi (n_a-n_b)} e^{+\ii\xi(\epsilon_a-\epsilon_b+\f{n}\cdot\f{\Omega})} e^{-\ii\phi \f{n}\cdot\f{\Omega}}\,,\nn
\gR^{ij,\f{n}} &\to& \gR^{ab,\f{n}} e^{-\ii\phi \f{n}\cdot\f{\Omega}}\,,
\end{eqnarray}
where $n_a\in\{0,1,2\}$ denotes the number of electrons in energy eigenstate $a$.
Thus, the Liouvillian is now dependent on the particle counting field $\chi$, the electronic energy counting field $\xi$, 
and the phonon energy counting field $\phi$, which we may for brevity combine in a vector $\f{\chi}=(\chi,\xi,\phi)$.
The characteristic polynomial $\mathcal{D}(\f{\chi}) = \abs{\mathcal{L}(\f{\chi})-\lambda\f{1}}$ 
of the now counting-field dependent Liouvillian formally equates to
\begin{eqnarray}
\label{eq:charPol}
\mathcal{D}&=&[\mathcal{L}_{11}-\lambda][\mathcal{L}_{22}-\lambda][\mathcal{L}_{33}-\lambda][\mathcal{L}_{44}-\lambda]\nn
&&-[\mathcal{L}_{11}-\lambda][\mathcal{L}_{22}-\lambda]\mathcal{L}_{34}\mathcal{L}_{43}\nn
&&-[\mathcal{L}_{11}-\lambda][\mathcal{L}_{33}-\lambda]\mathcal{L}_{24}\mathcal{L}_{42}\nn
&&-[\mathcal{L}_{11}-\lambda][\mathcal{L}_{44}-\lambda]\mathcal{L}_{23}\mathcal{L}_{32}\nn
&&-[\mathcal{L}_{22}-\lambda][\mathcal{L}_{44}-\lambda]\mathcal{L}_{13}\mathcal{L}_{31}\nn
&&-[\mathcal{L}_{33}-\lambda][\mathcal{L}_{44}-\lambda]\mathcal{L}_{12}\mathcal{L}_{21}\nn
&&+[\mathcal{L}_{11}-\lambda]\left[\mathcal{L}_{23}\mathcal{L}_{34}\mathcal{L}_{42}+\mathcal{L}_{24}\mathcal{L}_{43}\mathcal{L}_{32}\right]\nn
&&+[\mathcal{L}_{44}-\lambda]\left[\mathcal{L}_{12}\mathcal{L}_{23}\mathcal{L}_{31}+\mathcal{L}_{13}\mathcal{L}_{32}\mathcal{L}_{21}\right]\nn
&&+\mathcal{L}_{12}\mathcal{L}_{21}\mathcal{L}_{34}\mathcal{L}_{43}+\mathcal{L}_{13}\mathcal{L}_{31}\mathcal{L}_{24}\mathcal{L}_{42}\nn
&&-\mathcal{L}_{12}\mathcal{L}_{24}\mathcal{L}_{43}\mathcal{L}_{31}-\mathcal{L}_{13}\mathcal{L}_{34}\mathcal{L}_{42}\mathcal{L}_{21}\,,
\end{eqnarray}
where it should be kept in mind that the counting fields only occur in the off-diagonal ($\mathcal{L}_{i \neq j}$) contributions.
With the relations ($\sigma\in\{\rm L,R\}$)
\begin{eqnarray}\label{eq:local_detailed_balance}
\frac{\Gamma_{\rm ph}^{-+}}{\Gamma_{\rm ph}^{+-}} &=& e^{+\beta_{\rm ph}(\eP-\eM)}\,,\nn
\frac{\Gamma_\sigma^{0-,-\f{n}}}{\Gamma_\sigma^{-0,+\f{n}}} &=& e^{+\beta_\sigma(\eM-\eO-\mu_\sigma+\f{n}\cdot\f{\Omega})} e^{-\beta_{\rm ph} \f{n}\cdot\f{\Omega}}\,,\nn
\frac{\Gamma_\sigma^{0+,-\f{n}}}{\Gamma_\sigma^{+0,+\f{n}}} &=& e^{+\beta_\sigma(\eP-\eO-\mu_\sigma+\f{n}\cdot\f{\Omega})} e^{-\beta_{\rm ph} \f{n}\cdot\f{\Omega}}\,,\nn
\frac{\Gamma_\sigma^{-2,-\f{n}}}{\Gamma_\sigma^{2-,+\f{n}}} &=& e^{+\beta_\sigma(\ePM-\eM-\mu_\sigma+\f{n}\cdot\f{\Omega})} e^{-\beta_{\rm ph} \f{n}\cdot\f{\Omega}}\,,\nn
\frac{\Gamma_\sigma^{+2,-\f{n}}}{\Gamma_\sigma^{2+,+\f{n}}} &=& e^{+\beta_\sigma(\ePM-\eP-\mu_\sigma+\f{n}\cdot\f{\Omega})} e^{-\beta_{\rm ph} \f{n}\cdot\f{\Omega}}
\end{eqnarray}
one can show (compare Appendix~\ref{App:symmetries}) that the characteristic polynomial stays invariant under the replacements
\begin{eqnarray}
\label{eq:symmetry}
-\chi&\rightarrow&+\chi+\ii(\bL\mL-\bR\mR)\,,\nn
-\xi&\rightarrow&+\xi+\ii(\bR-\bL)\,,\\
-\phi&\rightarrow&+\phi+\ii(\bR-\bPH)\nonumber\,,
\end{eqnarray}
where we recover the affinities in Eq.~(\ref{EQ:entprodfull}).
This symmetry transfers to the long-term cumulant-generating function, and thus, 
the steady state fluctuation theorem for entropy production reads
\begin{eqnarray}
\lim_{t\to\infty}\frac{P_{+n_{\rm L}, +e_{\rm L}, +e_{\rm ph}}(t)}{P_{-n_{\rm L}, -e_{\rm L}, -e_{\rm ph}}(t)}=e^{\bf n \boldsymbol{\Delta}}\,,
\end{eqnarray}
with $\f n\equiv(n_{\rm L}, e_{\rm L}, e_{\rm ph})^{\rm T}$ and $\boldsymbol{\Delta}=(\bL\mL-\bR\mR,\bR-\bL,\bR-\bPH)^{\rm T}$.
Due to the similar three-terminal setup, the same fluctuation theorem can be obtained for the single electron transistor~\cite{schaller2013a}.

%%%%%%%%%%%%%%%%%%%%%%%%%%%%%%%%%%%%%%%%%%%%%%%%%%%%%%%%%%%%%%%%%%%%%%%%%%%%%%%
%%%%%%%%%%%%%%%%%%%%%%%%%%%%%%%%%%%%%%%%%%%%%%%%%%%%%%%%%%%%%%%%%%%%%%%%%%%%%%%
%%%%%%%%%%%%%%%%%%%%%%%%%%%%%%%%%%%%%%%%%%%%%%%%%%%%%%%%%%%%%%%%%%%%%%%%%%%%%%%
%%%%%%%%%%%%%%%%%%%%%%%%%%%%%%%%%%%%%%%%%%%%%%%%%%%%%%%%%%%%%%%%%%%%%%%%%%%%%%%

\section{Results}\label{sec:Results}

The implications of the resulting master equation are of course manifold. 
Below, we present a selection of the most interesting phonon-induced features.
For simplicity, we will discuss the case of symmetric couplings $\hqL=\hqR$ here.

%%%%%%%%%%%%%%%%%%%%%%%%%%%%%%%%%%%%%%%%%%%%%%%%%%%%%%%%%%%%%%%%%%%%%%%%%%%%%%%
%%%%%%%%%%%%%%%%%%%%%%%%%%%%%%%%%%%%%%%%%%%%%%%%%%%%%%%%%%%%%%%%%%%%%%%%%%%%%%%

\subsection{Electronic current versus internal bias}\label{sec:StationaryCurrent}

\begin{figure}[htb]
\begin{center}
\includegraphics[width=0.48\textwidth,clip=true]{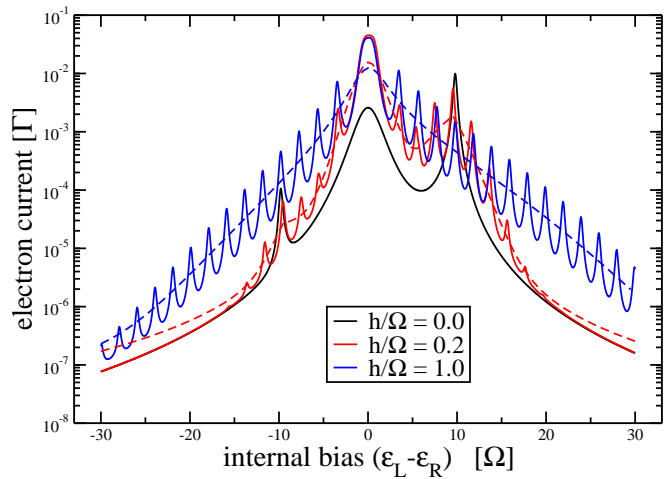}
\caption[]{\label{fig:CurrentDiffCoupl}
Electron current in units of $\gL=\gR=\Gamma$ versus the internal bias $\dE=\eL-\eR$ in units of $\Omega$. 
All graphs are evaluated far from the electronic wideband limit $\delta_{\rm L}/\Omega=\delta_{\rm R}/\Omega=\delta/\Omega=0.1$.
The electron-phonon couplings left and right are chosen equal $\hqL/\Omega=\hqR/\Omega=h/\Omega$.
The black line shows pure electronic transport decoupled from the phonon bath, $h/\Omega=0$. 
Due to the sharp Lorentzian shaped electronic tunneling rates observe two prominent electronic resonances.
When adding coupling to a single phonon mode (solid curves) we see that additional resonances appear.  
Caused by the on-site level configuration and large phonon bath temperatures ($\bPH\Omega=0.1$) the resonances approximately
symmetric in $\dE$.
At strong electron-phonon coupling resonances appear over the whole internal bias range (blue line).
This is different for coupling to a continuum phonon reservoir (dashed curves in background), 
where no additional resonances are found.
Other parameters are chosen as: $\Gamma/\Omega=0.01\,, U/\Omega=5.0\,,\tc/\Omega=1.0\,,\Phi=0\,,\beta_{\rm el} = \beta_{\rm ph}$
(implying $\Gamma \beta_{\rm el} = 10^{-3}$ and $\eP-\eM \ge 2 \abs{\tc}$).
Continuum parameters have been adjusted such that $\int_0^\infty J_\sigma(\omega)d\omega = \abs{h}^2$ and 
$\int_0^\infty J_\sigma(\omega)/\omega d\omega = \abs{h}^2/\Omega$.
}
\end{center}
\end{figure}

We compute the electronic matter current for coupling to a single phonon mode at frequency $\Omega$ and 
also for coupling to a continuum of phonons.
Fig.~\ref{fig:CurrentDiffCoupl} shows the electronic current at infinite external bias ($f_{\rm L}\to 1\,, f_{\rm R}\to 0$) but finite bandwidths as a 
function of the internal bias $\dE\equiv\eL-\eR$, which we define symmetrically with $\eL\equiv+\dE/2$ and $\eR\equiv-\dE/2$.
We note that due to the finite $\tc$, the system spectrum remains in the non-degenerate regime also when $\eL=\eR$.
The study of such currents is very common in theoretical~\cite{brandes2003a,brandes2004a,wang2012b,santamore2013a} studies as they reveal many internal details of the transport setup.
In Fig.~\ref{fig:CurrentDiffCoupl}, the black curve shows the 
pure electronic current without phonons ($\hqL=\hqR=0$) far away from the wide-band limit ($\delta_{\rm L}/\Omega=\delta_{\rm L}/\Omega=0.1$). 
Here, two electronic resonances at $\pm(\ePM-\eM)/\Omega=\pm10$ become visible. 
The Lorentzian shape of the graph is characteristic for such models and stems from the matrix elements in front of the rates.
For the colored  curves we increase the electron-phonon coupling ($\hqL=\hqR=h_q$) at large phonon bath temperature $\bPH\Omega=0.1$ (due
to the infinite-bias assumption the electronic temperature does not enter).
Due to the coupling to a single phonon mode we see additional side peaks appearing at $\dE=2n\Omega$ with integer $n$ (see solid red and blue curves), 
and these completely dominate the electronic peaks in the strong-coupling limit (solid blue).
For smaller phonon bath temperatures, the resonances would be more pronounced for positive $\dE$, since phonon emission into the bath is more likely (not shown).
Phonon induced oscillations in the electronic current as a function of the level detuning have been seen in experiments with InAs and graphene double quantum dots~\cite{roulleau2011}.
When we couple electronic transport to a continuum of phonon modes these detailed oscillations can not be resolved anymore (dashed curves in the background, see also 
the figure caption).

%%%%%%%%%%%%%%%%%%%%%%%%%%%%%%%%%%%%%%%%%%%%%%%%%%%%%%%%%%%%%%%%%%%%%%%%%%%%%%%
%%%%%%%%%%%%%%%%%%%%%%%%%%%%%%%%%%%%%%%%%%%%%%%%%%%%%%%%%%%%%%%%%%%%%%%%%%%%%%%

\subsection{Current/Dephasing rate versus external bias}\label{sec:DephasingRate}

\begin{figure}[htb]
\begin{center}
\includegraphics[width=0.48\textwidth,clip=true]{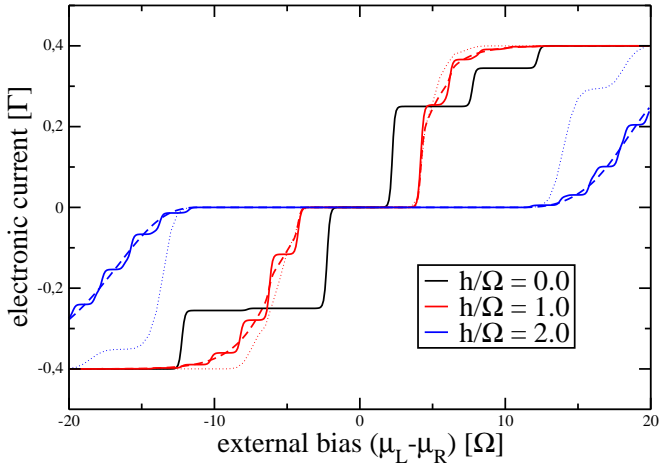}
\end{center}
\caption{\label{fig:Cur_h}
Plot of the electronic current versus the external bias voltage for different electron-phonon coupling strengths.
With increasing coupling strength, the steps corresponding to the bare electronic transitions (black curve) become supplemented
by additional plateaus accounting for an increasing number of phonons involved in the transport process.
The width of these smaller steps allows to determine the phonon frequency.
Consistently, the continuum phonon reservoir (dashed curves in background) does not exhibit these smaller steps.
Other parameters are chosen as: $\Gamma/\Omega=0.01\,,\tc/\Omega=1.0\,, \bL\Omega=\bR\Omega=\bPH\Omega=20.0\,,\delta_{\rm L}/\Omega=\delta_{\rm
R}/\Omega\to\infty\,,\eL/\Omega=-\eR/\Omega=0.5\,,U/\Omega=5.0\,,\Phi=0.0$ 
(implying $\Gamma \beta_{\rm el} = 0.2$ and $\eP-\eM = \sqrt{5} \abs{\tc}$).
Continuum parameters were adjusted such that $\int_0^\infty J_\sigma(\omega) d\omega = \abs{h}^2$ and 
$\int_0^\infty J_\sigma(\omega)/\omega d\omega = \abs{h}^2/\Omega$ (dashed curves).
Further approximating the continuum phonon correlation function with a single Gaussian as described in Sec.~\ref{sec:phonon_numeric}
yields for small bias quite analogous results (dotted curves).
}
\end{figure}

Typically, the current as a function of the external bias can be used to obtain internal system parameters via transport spectroscopy:
Transition frequencies of the system entering the transport window will -- at sufficiently small temperatures -- induce steps
in the current.
In Fig.~\ref{fig:Cur_h} we display the electronic matter current for different electron-phonon coupling strengths.
Whereas -- as a consequence of the phonon presence -- the single-mode version (solid curves) displays now many additional 
plateaus that allow e.g.\  for spectroscopy of the phonon frequency, 
the continuous phonon versions (dashed and dotted) only display a suppression of the current for small bias.
This phenomenon -- termed Franck-Condon blockade~\cite{koch2005a} -- is also observed when the phonons are taken into account dynamically.

Computing the dynamics of the coherences $\bra{-}\fRS(t)\ket{+}=(\bra{+}\fRS(t)\ket{-})^\ast$ yields a simple time evolution 
$\dot \fR_{-+}(t)=-\gamma \fR_{-+}(t)$, see Eq.~(\ref{EQ:coherenceevolution}).
This implies that the absolute square of $\fR_{-+}(t)$ decays exponentially with $\abs{\fR_{-+ }(t)}^2=e^{-2\Re(\gamma)t}\abs{\fR_{-+}(0)}^2$, 
where this dephasing is induced by both electronic and phononic reservoirs.
The dephasing rate $2\Re(\gamma)$ is a measure for the decay of the 
superposition of the states $\ket{-}$ and $\ket{+}$ to a classical mixture. 
When we neglect the asymmetry of the coupling $h_{qL}=h_{qR}$, the phonon correlation functions for the internal jumps vanish, and in consequence
also the internal transition rates $\gamma_{-+,-+}$ and $\gamma_{+-,+-}$ in Eq.~(\ref{EQ:coherenceevolution}) vanish.
Since furthermore diagonal matrix elements of the first four system coupling operators (such as e.g. $\bra{-} d_L \ket{-}$) vanish throughout, 
it also follows that $\gamma_{--,++}=0$, and we obtain for the dephasing rate 
\begin{eqnarray}
\label{eq:DephRate}
\mathcal{R}&=&[\Gamma^{0-}_{\rm L}+\Gamma^{0-}_{\rm R}+\Gamma^{0+}_{\rm L}+\Gamma^{0+}_{\rm R}\nn
&&+\Gamma^{2-}_{\rm L}+\Gamma^{2-}_{\rm R}+\Gamma^{2+}_{\rm L}+\Gamma^{2+}_{\rm R}]\,,
\end{eqnarray}
where we have used the abbreviations defined in Eq.~(\ref{EQ:defrateel}).
\begin{figure}[htb]
\begin{center}
\includegraphics[width=0.45\textwidth,clip=true]{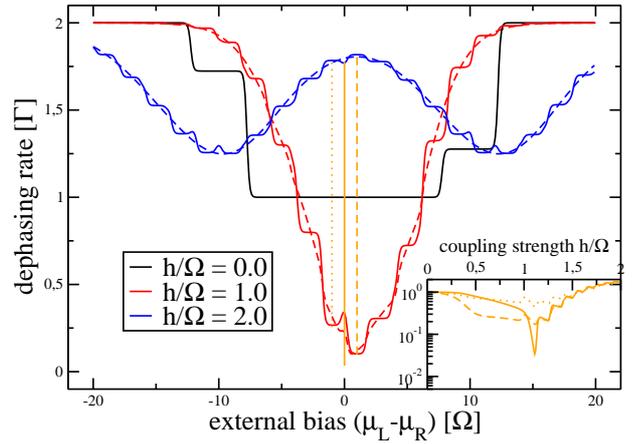}
\caption[]{\label{fig:RateBiasV}
Dephasing rate $\mathcal{R}$ in units of $\gL=\gR=\Gamma$ versus the external bias voltage $V$ in units of $\Omega$. 
The black reference curve shows the dephasing rate for pure electronic transport, $\hqL=\hqR=h=0.0$. 
Without phonon coupling, the dephasing rate where transport is dominated by 
the transitions $\ket{0}\to\ket{-},\ket{+}$ lies on the same level as the equilibrium dephasing rate, such that no step is visible.
If we increase the electron-phonon coupling $\hqL=\hqR$ we see a severe modulation of the curves. 
In the Franck-Condon regime around $V=0.0$ (vertical orange lines mark maximum and minimum dephasing rates in the 
interval $h/\Omega\in[0,2]$) the dephasing rate becomes suppressed for intermediate
electron-phonon coupling strengths (see the inset for the dephasing rate at $V\in\{-\Omega,0,+\Omega\}$).
Other parameters are chosen as: $\Gamma/\Omega=0.01\,,\tc/\Omega=1.0\,, \bL\Omega=\bR\Omega=\bPH\Omega=20.0\,,\delta_{\rm L}/\Omega=\delta_{\rm
R}/\Omega\to\infty\,,\eL/\Omega=-\eR/\Omega=0.5\,,U/\Omega=5.0\,,\Phi=0.0$ (implying $\Gamma \beta_{\rm el} = 0.2$ 
and $\eP-\eM=\sqrt{5}\abs{\tc}$).
Continuum parameters were adjusted such that $\int_0^\infty J_\sigma(\omega) d\omega = \abs{h}^2$ and 
$\int_0^\infty J_\sigma(\omega)/\omega d\omega = \abs{h}^2/\Omega$.
}
\end{center}
\end{figure}
The phonon plateaus are also very well visible in the dephasing rate, see Fig.~\ref{fig:RateBiasV}.
Counter-intuitively, when we increase the electron-phonon coupling the dephasing rate first decreases before it increases again (compare orange curves in the inset).
This suppression occurs in the current blockade regime.
Interestingly, the dephasing rate becomes much smaller than the equilibrium dephasing rate observed without phonons.
Thus, we find that while increasing the coupling strength to the phonon reservoir, the model effectively shows a decrease of
the dephasing rate which is in stark contrast to general expectations.
We attribute this behaviour to the conditioned state of the phonon reservoir.
A more intuitive explanation is that the Franck-Condon blockade prevents transport through the
charge qubit and thereby also transport-associated decoherence.

%%%%%%%%%%%%%%%%%%%%%%%%%%%%%%%%%%%%%%%%%%%%%%%%%%%%%%%%%%%%%%%%%%%%%%%%%%%%%%%
%%%%%%%%%%%%%%%%%%%%%%%%%%%%%%%%%%%%%%%%%%%%%%%%%%%%%%%%%%%%%%%%%%%%%%%%%%%%%%%

\subsection{Thermoelectric Generator}

\begin{figure}[ht]
\includegraphics[width=0.48\textwidth,clip=true]{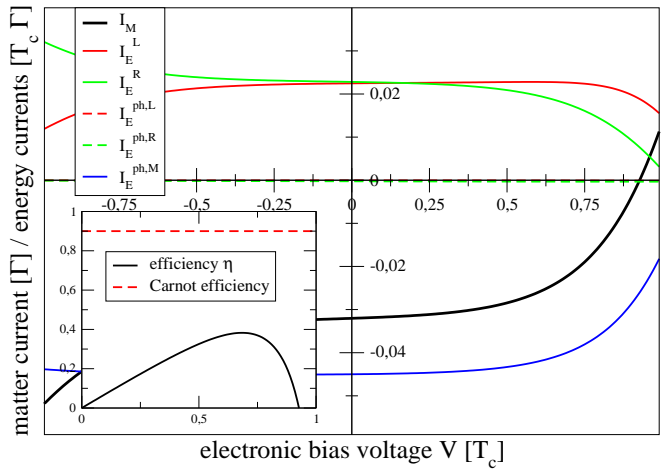}
\caption{\label{FIG:heatcurrents}(Color Online)
Plot of the matter and energy currents for a hot phonon and cold electronic reservoirs versus electronic bias
voltage.
In the lower right quadrant, the electronic matter current (bold black) runs against a potential gradient thereby
generating power $P_{\rm out}=-I_{\rm M} V$.
The first law manifests in the fact that all energy currents add up to zero.
Parameters have been chosen such that the internal phonon-assisted transitions between eigenstates $\ket{-}$ and
$\ket{+}$ dominate the phonon heat flow (solid blue versus dashed curves for external jumps).
Relating the power output with the heat input from the phonon reservoir $Q = +(I_E^{\rm ph,L}+I_E^{\rm ph,R}+I_{\rm E}^{\rm ph,M})$
we see that the efficiency of this process (inset, for positive bias voltage only) remains significantly below Carnot efficiency.
Other parameters: $\gL=\gR=\Gamma = 0.01 \tc$, $J_{\rm L} T_{\rm c}^2 =J_{\rm R} T_{\rm c}^2=0.001$, $J_0 T_{\rm c}^2 =1.0$, 
$w_{\rm c}^{\rm L}=w_{\rm c}^{\rm R} = w_{\rm c} = 1.0 T_{\rm c}$, 
$\upvarepsilon_{\rm L}=+0.5 T_{\rm c}=-\upvarepsilon_{\rm R}$, $U=5.0 T_{\rm c}$, $\bL T_{\rm c}=\bR T_{\rm c}=10.0$, $\beta_{\rm ph} T_{\rm c} = 1.0$
(implying $\beta_{\rm el} \Gamma = 0.1$ and $\eP-\eM=\sqrt{5}\abs{\tc}$)
}
\end{figure}

Multi-terminal nanostructures may serve as nanomachines converting e.g.\  temperature gradients into 
electric power.
Here, we consider the case where a hot phonon bath and cold electronic reservoirs 
may induce an electronic current at vanishing bias -- or even a current against a finite
bias generating useful power.
We note that whereas for a single-electron transistor (with its always-symmetric electron-phonon coupling)
one would require non-flat electronic tunneling rates to see such an effect, this is different in the
present model when we apply it to the case of a continuous phonon spectrum.
Formally, we consider in Eq.~(\ref{EQ:entprodred}) a situation where the matter current $I_{\rm M}=I_{\rm M}^{\rm L}$ from left to
right is negative although $\mL<\mR$.
This is for $\beta_{\rm el} < \beta_{\rm ph}$ only possible when heat flows out of the hot phonon reservoir, with use
of Eq.~(\ref{EQ:entprodred}) more precisely when 
$I_{\rm E}^{\rm ph} \ge -\frac{\beta_{\rm el}}{\beta_{\rm el}-\beta_{\rm ph}} (\mu_L-\mu_R) I_{\rm M}^{\rm L} > 0$.
To quantify the performance of such a device, it is instructive to relate the power output  $P_{\rm out} = -I_{\rm M}^{\rm L}(\mL-\mR)=-I_{\rm M} V$
to the heat entering from the hot phonon reservoir $Q = I_{\rm E}^{\rm ph}$.
Positivity of the entropy production~(\ref{EQ:entprodred}) then grants that the efficiency of this process
\begin{eqnarray}
\eta = \frac{P_{\rm out}}{Q} = - \frac{I_{\rm M} V}{I_{\rm E}^{\rm ph}} \le 1 - \frac{T_{\rm el}}{T_{\rm ph}} = \eta_{\rm Ca}
\end{eqnarray}
is upper-bounded by Carnot efficiency.
In general however, the efficiency can be significantly smaller as is illustrated in Fig.~\ref{FIG:heatcurrents}.
In fact, the inset shows that Carnot efficiency is not even reached at the new equilibrium, where the electronic 
matter current vanishes.
Formally, this is due to the fact that -- in contrast to previous weak-coupling models~\cite{krause2011a,schaller2014} -- 
the total entropy production does not vanish at this point.
This is somewhat expected, since due to the presence of phonons, our model does not obey the the tight-coupling condition~\cite{gomez_marin2006a}.

%%%%%%%%%%%%%%%%%%%%%%%%%%%%%%%%%%%%%%%%%%%%%%%%%%%%%%%%%%%%%%%%%%%%%%%%%%%%%%%
%%%%%%%%%%%%%%%%%%%%%%%%%%%%%%%%%%%%%%%%%%%%%%%%%%%%%%%%%%%%%%%%%%%%%%%%%%%%%%%
%%%%%%%%%%%%%%%%%%%%%%%%%%%%%%%%%%%%%%%%%%%%%%%%%%%%%%%%%%%%%%%%%%%%%%%%%%%%%%%
%%%%%%%%%%%%%%%%%%%%%%%%%%%%%%%%%%%%%%%%%%%%%%%%%%%%%%%%%%%%%%%%%%%%%%%%%%%%%%%

\section{Summary}\label{sec:Summary}

We have investigated coherent electronic transport strongly coupled to vibrations.
To obtain a thermodynamically consistent master equation, the secular-approximation has to be performed
in the new system basis that arises after the polaron transform.
The method presented here yields a low dimensional master equation in Lindblad form which 
accounts for thermodynamic consistency although in the original frame the phonons are in a displaced 
thermal state.
Thermodynamic consistency has been demonstrated by an analytic proof of the 
fluctuation theorem for entropy production.

Using the Full Counting Statistics we computed the electronic current versus internal and external bias
and reproduced electron-phonon-induced features such as oscillations versus the internal bias and
signatures of Franck-Condon blockade.
We stress that the description of this rich dynamics required only the four states of the double quantum
dot to be taken into account dynamically.

We have found that the dephasing rate of coherences in the pointer basis behaves in some regimes counter-intuitively
as a function of the electron-phonon coupling strength.
A simple intuitive explanation for this behaviour is that the Franck-Condon blockade stabilizes coherences, 
thereby also blocking transport through the DQD circuit.
 
The analysis of the entropy production in the polaron master equation has allowed us to study the performance of the
system when viewed as a thermoelectric generator converting a temperature gradient into electric power.
We have found that in the strong-coupling regime, the system deviates strongly from tight-coupling between energy
and matter current, and consequently, the efficiency for this process was found to be significantly below Carnot efficiency.
 
Finally, we want to mention that our method can be generalized to more complex systems and thus allows 
applications in a variety of transport setups involving phonons such as molecules.
%

%%%%%%%%%%%%%%%%%%%%%%%%%%%%%%%%%%%%%%%%%%%%%%%%%%%%%%%%%%%%%%%%%%%%%%%%%%%%%%%%
%%%%%%%%%%%%%%%%%%%%%%%%%%%%%%%%%%%%%%%%%%%%%%%%%%%%%%%%%%%%%%%%%%%%%%%%%%%%%%%%
%%%%%%%%%%%%%%%%%%%%%%%%%%%%%%%%%%%%%%%%%%%%%%%%%%%%%%%%%%%%%%%%%%%%%%%%%%%%%%%%
%%%%%%%%%%%%%%%%%%%%%%%%%%%%%%%%%%%%%%%%%%%%%%%%%%%%%%%%%%%%%%%%%%%%%%%%%%%%%%%%

\section{Acknowledgments}

T.~B. and G.~S. gratefully acknowledge financial support
by the DFG (SFB 910, GRK 1588, SCHA \mbox{1642/2-1}).
M.~E. has been supported by the National Research Fund, Luxembourg, 
in the frame of project \mbox{FNR/A11/02}.

%%%%%%%%%%%%%%%%%%%%%%%%%%%%%%%%%%%%%%%%%%%%%%%%%%%%%%%%%%%%%%%%%%%%%%%%%%%%%%%%
%%%%%%%%%%%%%%%%%%%%%%%%%%%%%%%%%%%%%%%%%%%%%%%%%%%%%%%%%%%%%%%%%%%%%%%%%%%%%%%%
%%%%%%%%%%%%%%%%%%%%%%%%%%%%%%%%%%%%%%%%%%%%%%%%%%%%%%%%%%%%%%%%%%%%%%%%%%%%%%%%
%%%%%%%%%%%%%%%%%%%%%%%%%%%%%%%%%%%%%%%%%%%%%%%%%%%%%%%%%%%%%%%%%%%%%%%%%%%%%%%%

\bibliographystyle{unsrt}
\bibliography{literatur_edit}

%%%%%%%%%%%%%%%%%%%%%%%%%%%%%%%%%%%%%%%%%%%%%%%%%%%%%%%%%%%%%%%%%%%%%%%%%%%%%%%%
%%%%%%%%%%%%%%%%%%%%%%%%%%%%%%%%%%%%%%%%%%%%%%%%%%%%%%%%%%%%%%%%%%%%%%%%%%%%%%%%
%%%%%%%%%%%%%%%%%%%%%%%%%%%%%%%%%%%%%%%%%%%%%%%%%%%%%%%%%%%%%%%%%%%%%%%%%%%%%%%%
%%%%%%%%%%%%%%%%%%%%%%%%%%%%%%%%%%%%%%%%%%%%%%%%%%%%%%%%%%%%%%%%%%%%%%%%%%%%%%%%

\appendix

\section{Polaron transformation}\label{App:PolaronTrafo}
We consider the polaron transformation
\begin{eqnarray}
U=e^{\dL^\dagger \dL\mathcal{B}_{\rm L}+\dR^\dagger \dR \mathcal{B}_{\rm R}}\,,
\end{eqnarray}
with the fermionic annihilation operators $d_\sigma$ and the bosonic operators
\begin{eqnarray}
\mathcal{B}_\sigma=\sum_q\left(\frac{h^\ast_{q,\sigma}}{\omega_q}a_q^\dagger-\frac{h_{q,\sigma}}{\omega_q}a_q\right)
\end{eqnarray}
with bosonic annihilation operators $a_q$. 
To calculate the transformation rules, we recall the BCH relation
\begin{eqnarray}
e^X Y e^{-X}&=&\sum^\infty_{n=0}\frac{1}{n!}\com{X,Y}_n\,,
\end{eqnarray}
with the short-hand notation
$\com{X,Y}_{n+1}=\com{X,\com{X,Y}_n}$ and $\com{X,Y}_0=Y$.
We first note that the exponential in the polaron transformation can be written in a separated fashion
\begin{eqnarray}
U&=&e^{\dL^\dagger\dL\mathcal{B}_{\rm L}}e^{\dR^\dagger\dR\mathcal{B}_{\rm R}}e^{-\dL^\dagger\dL\dR^\dagger\dR\com{\mathcal{B}_{\rm L},\mathcal{B}_{\rm R}}/2}\nn
&\equiv&U_{\rm L}U_{\rm R}U_{\rm LR}\,,\nn
U_{\rm LR}&=&e^{\dL^\dagger\dL\dR^\dagger\dR \ii\Phi}\,,\nn
\ii\Phi&\equiv&\com{\mathcal{B}_{\rm R},\mathcal{B}_{\rm L}}/2\,,
\end{eqnarray}
where it is easy to show that $\Phi^\ast=\Phi$.
Consequently, the adjoint operator is given by
\begin{eqnarray}
U^\dagger=U^\dagger_{\rm LR}U^\dagger_{\rm R}U^\dagger_{\rm L}\,,
\end{eqnarray}
and we note that $\com{U_{\rm LR},U_{\rm L}}=\com{U_{\rm LR},U_{\rm R}}=0$.
Alternatively, we can also split the unitary transformation according to
\begin{eqnarray}
U=U_{\rm R}U_{\rm L}U^\dagger_{\rm LR}\,,\quad U^\dagger=U_{\rm LR}U_{\rm L}^\dagger U_{\rm R}^\dagger\,,
\end{eqnarray}
where again $\com{U_{\rm LR},U_{\rm L}^\dagger}=\com{U_{\rm LR},U_{\rm R}^\dagger}=0$ holds.

%%%%%%%%%%%%%%%%%%%%%%%%%%%%%%%%%%%%%%%%%%%%%%%%%%%%%%%%%%%%%%%%%%%%%%%%%%%%%%%%
%%%%%%%%%%%%%%%%%%%%%%%%%%%%%%%%%%%%%%%%%%%%%%%%%%%%%%%%%%%%%%%%%%%%%%%%%%%%%%%%

\subsection{Left Mode Operators}
We consider the action of the Polaron transformation on the left fermionic annihilation operator
\begin{eqnarray}
U\dL U^\dagger&=&U_{\rm L}U_{\rm R}U_{\rm LR}\dL U^\dagger_{\rm LR}U^\dagger_{\rm R}U^\dagger_{\rm L}=U_{\rm L}U_{\rm LR}\dL U^\dagger_{\rm LR}U^\dagger_{\rm L}\nn
&=&U_{\rm L}\dL e^{-\dR^\dagger\dR\ii\Phi} U^\dagger_{\rm L}=U_{\rm L}\dL U^\dagger_{\rm L} e^{-\dR^\dagger\dR\ii\Phi}\nn
&=&\dL e^{-\mathcal{B}_{\rm L}}e^{-\dR^\dagger\dR\ii\Phi} \,.
\end{eqnarray}
The left fermionic creation operator then transforms according to
\begin{eqnarray}
U\dL^\dagger U^\dagger=\dL^\dagger e^{+\mathcal{B}_{\rm L}}e^{+\dR^\dagger\dR\ii\Phi} \,.
\end{eqnarray}

%%%%%%%%%%%%%%%%%%%%%%%%%%%%%%%%%%%%%%%%%%%%%%%%%%%%%%%%%%%%%%%%%%%%%%%%%%%%%%%%
%%%%%%%%%%%%%%%%%%%%%%%%%%%%%%%%%%%%%%%%%%%%%%%%%%%%%%%%%%%%%%%%%%%%%%%%%%%%%%%%

\subsection{Right Mode Operators}
In a similar fashion, we evaluate the transformation of the right fermionic annihilation operator
\begin{eqnarray}
U\dR U^\dagger&=&U_{\rm R}U_{\rm L}U_{\rm LR}^\dagger \dR U_{\rm LR}U^\dagger_{\rm L}U^\dagger_{\rm R}=U_{\rm R}U_{\rm LR}^\dagger\dR U_{\rm LR}U^\dagger_{\rm R}\nn
&=&U_{\rm R}\dR e^{+\dL^\dagger\dL\ii\Phi} U^\dagger_{\rm R}=U_{\rm R}\dR U^\dagger_{\rm R} e^{+\dL^\dagger\dL\ii\Phi}\nn
&=&\dR e^{-\mathcal{B}_{\rm R}}e^{+\dL^\dagger\dL\ii\Phi} \,,
\end{eqnarray}
and the adjoint operator becomes
\begin{eqnarray}
U\dR^\dagger U^\dagger=\dR^\dagger e^{+\mathcal{B}_{\rm R}}e^{-\dL^\dagger\dL\ii\Phi} \,.
\end{eqnarray}

%%%%%%%%%%%%%%%%%%%%%%%%%%%%%%%%%%%%%%%%%%%%%%%%%%%%%%%%%%%%%%%%%%%%%%%%%%%%%%%%
%%%%%%%%%%%%%%%%%%%%%%%%%%%%%%%%%%%%%%%%%%%%%%%%%%%%%%%%%%%%%%%%%%%%%%%%%%%%%%%%

\subsection{Bosonic Operators}
For the bosonic annihilation operator we obtain
\begin{eqnarray} 
U a_q U^\dagger&=&U_{\rm L}U_{\rm R}U_{\rm LR}a_q U^\dagger_{\rm LR}U^\dagger_{\rm R}U^\dagger_{\rm L}=U_{\rm L}U_{\rm R}\dL U^\dagger_{\rm R}U^\dagger_{\rm L}\nn
&=&U_{\rm L}\left[a_q-\frac{h^\ast_{q,\rm R}}{\omega_q}\dR^\dagger\dR\right] U^\dagger_{\rm L}\nn
&=&U_{\rm L}a_q U^\dagger_{\rm L}-\frac{h^\ast_{q,\rm R}}{\omega_q}\dR^\dagger\dR \nn
&=&a_q -\frac{h^\ast_{q,\rm L}}{\omega_q}\dL^\dagger\dL-\frac{h^\ast_{q,\rm R}}{\omega_q}\dR^\dagger\dR
\end{eqnarray}
and similarly for the creation operator
\begin{eqnarray}
U a_q^\dagger U^\dagger=a_q^\dagger -\frac{h_{q,\rm L}}{\omega_q}\dL^\dagger\dL-\frac{h_{q,\rm R}}{\omega_q}\dR^\dagger\dR \,.
\end{eqnarray}

%%%%%%%%%%%%%%%%%%%%%%%%%%%%%%%%%%%%%%%%%%%%%%%%%%%%%%%%%%%%%%%%%%%%%%%%%%%%%%%%
%%%%%%%%%%%%%%%%%%%%%%%%%%%%%%%%%%%%%%%%%%%%%%%%%%%%%%%%%%%%%%%%%%%%%%%%%%%%%%%%

\subsection{Polaron transformation of the DQD Hamiltonian}
The total Hamiltonian of the DQD is given by
\begin{eqnarray}
\mathcal{H} &=& \sum_{k\sigma} \upvarepsilon_{k\sigma} c_{k\sigma}^\dagger c_{k\sigma} + \sum_q \omega_q a_q^\dagger a_q\nn
&&+\eL \dL^\dagger \dL + \eR \dR^\dagger \dR + U \dL^\dagger \dL \dR^\dagger \dR\nn
&&+\tc (\dL \dR^\dagger + \dR \dL^\dagger)\nn
&&+\sum_{k\sigma} \left(t_{k\sigma} d_\sigma c_{k\sigma}^\dagger + {\rm h.c.}\right)\nn
&&+\sum_{q\sigma} \left(h_{q\sigma} a_q + h_{q\sigma}^* a_q^\dagger\right) d_\sigma^\dagger d_\sigma\,.
\end{eqnarray}
Applying the polaron transformation to the total Hamiltonian $\bar{\mathcal{H}}=U \mathcal{H} U^\dagger$ implies that some parts of the Hamiltonian will change.
In particular, we have for the free bosonic Hamiltonian
\begin{eqnarray}
H'_{\rm ph} &=& \sum_q \omega_q \left(a_q^\dagger - \frac{\hqL}{\omega_q} \dL^\dagger \dL - \frac{\hqR}{\omega_q} \dR^\dagger \dR\right)\times\nn
&&\times\left(a_q -\frac{\hqL^*}{\omega_q} \dL^\dagger \dL - \frac{\hqR^*}{\omega_q} \dR^\dagger \dR\right)\,,
\end{eqnarray}
for the electronic inter-dot tunneling Hamiltonian
\begin{eqnarray}
H'_T &=& \tc \dL e^{-(\dL^\dagger \dL + \dR^\dagger \dR)\ii\Phi} \dR^\dagger e^{-\mathcal{B}_{\rm L}} e^{+\mathcal{B}_{\rm R}}\nn
&&+ \tc \dR e^{+(\dL^\dagger \dL + \dR^\dagger \dR)\ii\Phi} \dL^\dagger e^{-\mathcal{B}_{\rm R}} e^{+\mathcal{B}_{\rm L}}\nn
&=& \tc e^{-2\ii\Phi} \dL \dR^\dagger e^{-\mathcal{B}_{\rm L}} e^{+\mathcal{B}_{\rm R}}\nn
&&+ \tc e^{+2\ii\Phi} \dR \dL^\dagger e^{-\mathcal{B}_{\rm R}} e^{+\mathcal{B}_{\rm L}}\,,
\end{eqnarray}
for the electron-lead tunneling Hamiltonian
\begin{eqnarray}
H'_V &=& \sum_k \left(\tkL \dL e^{-\dR^\dagger \dR\ii\Phi} e^{-\mathcal{B}_{\rm L}} c_{kL}^\dagger + {\rm h.c.}\right)\nn
&&+\sum_k \left(\tkR \dR e^{+\dL^\dagger \dL\ii\Phi} e^{-\mathcal{B}_{\rm R}} c_{kR}^\dagger + {\rm h.c.}\right)\,,
\end{eqnarray}
and for the electron-phonon interaction
\begin{eqnarray}
H'_{\rm e-ph} &=& \sum_{q\sigma} \left(h_{q\sigma} a_q + h_{q\sigma}^* a_q^\dagger\right) d_\sigma^\dagger d_\sigma\\
&&- \sum_{q\sigma} h_{q\sigma} 
\left(\frac{\hqL^*}{\omega_q} \dL^\dagger \dL + \frac{\hqR^*}{\omega_q} \dR^\dagger \dR\right) d_\sigma^\dagger d_\sigma\nn
&&- \sum_{q\sigma} h_{q\sigma}^* 
\left(\frac{\hqL}{\omega_q} \dL^\dagger \dL + \frac{\hqR}{\omega_q} \dR^\dagger \dR\right) d_\sigma^\dagger d_\sigma\,.
\nonumber
\end{eqnarray}

For the sum of the free phonon and the electron-phonon interaction Hamiltonians we obtain
\begin{eqnarray}
H'_{\rm ph} + H'_{\rm e-ph} &=& \sum_q \omega_q a_q^\dagger a_q\\
&& - \sum_q \left(\frac{\abs{\hqL}^2}{\omega_q} \dL^\dagger \dL + \frac{\abs{\hqR}^2}{\omega_q} \dR^\dagger \dR\right)\nn
&& - \sum_q \frac{\hqL^* \hqR + \hqL \hqR^*}{\omega_q} \dL^\dagger \dL \dR^\dagger \dR\,.\nonumber
\end{eqnarray}
Therefore, the total Hamiltonian after the polaron transformation reads
\begin{eqnarray}
H &=& \sum_{k\sigma} \upvarepsilon_{k\sigma} c_{k\sigma}^\dagger c_{k\sigma} + \sum_q \omega_q a_q^\dagger a_q\nn
&&+\etL \dL^\dagger \dL + \etR \dR^\dagger \dR + \bar U \dL^\dagger \dL \dR^\dagger \dR\nn
&&+\tc e^{-2\ii\Phi} \dL \dR^\dagger e^{-\mathcal{B}_{\rm L}} e^{+\mathcal{B}_{\rm R}}\nn
&&+\tc e^{+2\ii\Phi} \dR \dL^\dagger e^{-\mathcal{B}_{\rm R}} e^{+\mathcal{B}_{\rm L}}\nn
&&+\sum_k \left(\tkL \dL e^{-\dR^\dagger \dR\ii\Phi} e^{-\mathcal{B}_{\rm L}} c_{kL}^\dagger + {\rm h.c.}\right)\nn
&&+\sum_k \left(\tkR \dR e^{+\dL^\dagger \dL\ii\Phi} e^{-\mathcal{B}_{\rm R}} c_{kR}^\dagger + {\rm h.c.}\right)\,,
\end{eqnarray}
with renormalized on-site energies~(\ref{eq:RenormEnergy}) and the Coulomb interaction~(\ref{eq:RenormCBrep}).
When furthermore one demands that all expectation values of reservoir coupling operators should vanish (see below), 
one arrives at the splitting into system, reservoir, and interaction parts used in the paper.

%%%%%%%%%%%%%%%%%%%%%%%%%%%%%%%%%%%%%%%%%%%%%%%%%%%%%%%%%%%%%%%%%%%%%%%%%%%%%%%%
%%%%%%%%%%%%%%%%%%%%%%%%%%%%%%%%%%%%%%%%%%%%%%%%%%%%%%%%%%%%%%%%%%%%%%%%%%%%%%%%
%%%%%%%%%%%%%%%%%%%%%%%%%%%%%%%%%%%%%%%%%%%%%%%%%%%%%%%%%%%%%%%%%%%%%%%%%%%%%%%%
%%%%%%%%%%%%%%%%%%%%%%%%%%%%%%%%%%%%%%%%%%%%%%%%%%%%%%%%%%%%%%%%%%%%%%%%%%%%%%%%

\section{Shift factor}\label{App:ShiftFactor}
We use that for a thermal state $\rho \propto e^{-\beta\sum_q\omega_q a_q^\dagger a_q}$, one has
for all complex-valued numbers $\alpha_q$
\begin{eqnarray}
\expval{e^{-\sum_q(\alpha_q a^\dagger_q-\alpha^\ast_q a_q)}}=e^{-\sum_q\abs{\alpha_q}^2[n_{\rm B}(\omega_q)+1/2]}
\end{eqnarray}
with the Bose-distribution $n_{\rm B}(\omega_q)=[e^{\beta\omega_q}-1]^{-1}$.
Applying that to the shift factor $\kappa$, for generality in the interaction picture, we obtain
\begin{eqnarray}
\kappa&=&\expval{e^{-\boldsymbol{\mathcal{B}}_{\rm L}(\tau)}e^{+\boldsymbol{\mathcal{B}}_{\rm R}(\tau)}}=\expval{e^{-\boldsymbol{\mathcal{B}}_{\rm L}(\tau)+\boldsymbol{\mathcal{B}}_{\rm R}(\tau)}}e^{\ii\Phi}\nn
&=&\expval{e^{\sum_q\left(\frac{h_{q,\rm R}^\ast-h_{q,\rm L}^\ast}{\omega_q}a_q^\dagger e^{+\ii \omega_q\tau}-\frac{h_{q,\rm R}-h_{q,\rm L}}{\omega_q}a_q e^{-\ii \omega_q\tau}\right)}}e^{+\ii\Phi}\nn
&=&e^{-\sum_q\frac{\abs{h_{q,\rm R}-h_{q,\rm L}}^2}{\omega_q^2}[n_{\rm B}(\omega_q)+1/2]}e^{+\ii\Phi}\,,
\end{eqnarray}
and see that $\kappa$ is independent of $\tau$.
Therefore, we can already in the Schr\"odinger picture write the Hamiltonian in a way that is suitable for the derivation of a master equation
with splitting into system, reservoir and interaction parts given by
Eqns.~(\ref{EQ:hamsystem}),~(\ref{eq:HBpol}), and~(\ref{eq:HV}) and~(\ref{eq:HT}) in the paper, respectively.

%%%%%%%%%%%%%%%%%%%%%%%%%%%%%%%%%%%%%%%%%%%%%%%%%%%%%%%%%%%%%%%%%%%%%%%%%%%%%%%%
%%%%%%%%%%%%%%%%%%%%%%%%%%%%%%%%%%%%%%%%%%%%%%%%%%%%%%%%%%%%%%%%%%%%%%%%%%%%%%%%
%%%%%%%%%%%%%%%%%%%%%%%%%%%%%%%%%%%%%%%%%%%%%%%%%%%%%%%%%%%%%%%%%%%%%%%%%%%%%%%%
%%%%%%%%%%%%%%%%%%%%%%%%%%%%%%%%%%%%%%%%%%%%%%%%%%%%%%%%%%%%%%%%%%%%%%%%%%%%%%%%

\section{Inverse polaron transform}\label{App:polaron_inverse}

To apply the inverse polaron transformation, it is useful to write it conditioned on the electronic occupation
\begin{eqnarray}
U &=& \f{1} + \dL^\dagger \dL \left(e^{\mathcal{B}_{\rm L}}-\f{1}\right) + \dR^\dagger \dR \left(e^{\mathcal{B}_{\rm R}}-\f{1}\right)\nn
&&+ \dL^\dagger \dL  \dR^\dagger \dR \left(e^{\mathcal{B}_{\rm L}+\mathcal{B}_{\rm R}} -e^{\mathcal{B}_{\rm L}}-e^{\mathcal{B}_{\rm R}}+\f{1}\right)\nn
&=& P_0 \f{1} + P_{\rm L} e^{\mathcal{B}_{\rm L}} + P_{\rm R} e^{\mathcal{B}_{\rm R}} + P_2 e^{\mathcal{B}_{\rm L}+\mathcal{B}_{\rm R}}
\end{eqnarray}
where with the projectors $P_0= (\f{1}-\dL^\dagger \dL) (\f{1}-\dR^\dagger \dR)$, $P_2=\dL^\dagger \dL \dR^\dagger \dR$, $P_{\rm L} =  \dL^\dagger \dL (\f{1}-\dR^\dagger \dR)$, 
and $P_{\rm R} = (\f{1}-\dL^\dagger \dL) \dR^\dagger \dR$ it becomes visible that -- 
depending on the system state in the localized basis -- different unitary operations are 
applied on the reservoir.
For the phonon reservoir state this implies
\begin{eqnarray}
U^\dagger \RB^{\rm ph} U &=& P_0 \otimes  \RB^{\rm ph} + P_2 \otimes e^{-(\mathcal{B}_{\rm L}+\mathcal{B}_{\rm R})} \RB^{\rm ph} e^{+(\mathcal{B}_{\rm L}+\mathcal{B}_{\rm R})}\nn
&&+P_{\rm L} \otimes e^{-\mathcal{B}_{\rm L}} \RB^{\rm ph} e^{+\mathcal{B}_{\rm L}}+ P_{\rm R} \otimes e^{-\mathcal{B}_{\rm R}} \RB^{\rm ph} e^{+\mathcal{B}_{\rm R}}\,.\nn
\end{eqnarray}
Considering that these unitary operations displace the phonons
\begin{eqnarray}
e^{-\mathcal{B}_\sigma} a_q^\dagger a_q e^{+\mathcal{B}_\sigma} &=& \left(a_q^\dagger + \frac{h_{q\sigma}}{\omega_q}\right) \left(a_q + \frac{h_{q\sigma}^*}{\omega_q}\right)\,,\nn
e^{-\mathcal{B}_{\rm L}-\mathcal{B}_{\rm R}} a_q^\dagger a_q e^{+\mathcal{B}_{\rm L}+\mathcal{B}_{\rm R}} &=& \left(a_q^\dagger + \frac{\hqL+\hqR}{\omega_q}\right)\times\nn
&&\times \left(a_q + \frac{\hqL^*+\hqR^*}{\omega_q}\right)
\end{eqnarray}
the reservoir state becomes the displaced thermal state -- conditioned on the electronic occupation of the system.

Specifically, when in the localized basis 
the system density matrix is written as
\begin{eqnarray}
\RS &=& \rho_0 P_0 + \rho_2 P_2 + \rho_{\rm L} P_{\rm L} + \rho_{\rm R} P_{\rm R}\nn
&&+\rho_{\rm LR} P_{\rm LR} + \rho_{\rm RL} P_{\rm RL}
\end{eqnarray}
with $P_{\rm LR} = \ket{\rm L}\bra{\rm R}$ and $P_{\rm RL} = \ket{\rm R}\bra{\rm L}$,
it transforms according to
\begin{eqnarray}
U^\dagger \RS U &=& \rho_0 P_0 + \rho_2 P_2 +  \rho_{\rm L} P_{\rm L} + \rho_{\rm R} P_{\rm R}\nn
&&+\rho_{\rm LR} P_{\rm L} P_{\rm LR} P_{\rm R} e^{-\mathcal{B}_{\rm L}} e^{+\mathcal{B}_{\rm R}}\nn
&&+\rho_{\rm RL} P_{\rm R} P_{\rm RL} P_{\rm L} e^{-\mathcal{B}_{\rm R}} e^{+\mathcal{B}_{\rm L}}\,.
\end{eqnarray}
This implies that the total system-phonon density matrix in the original frame is given by
\begin{eqnarray}
\tilde{\rho} &=& U^\dagger \RS \otimes \f{1} U U^\dagger \f{1} \otimes \RB^{\rm ph} U\nn
&=& \rho_0 P_0 \otimes \RB^{\rm ph} + \rho_2 P_2 \otimes e^{-(\mathcal{B}_{\rm L}+\mathcal{B}_{\rm R})} \RB^{\rm ph} e^{+(\mathcal{B}_{\rm L}+\mathcal{B}_{\rm R})}\nn
&&+\rho_{\rm L} P_{\rm L} \otimes e^{-\mathcal{B}_{\rm L}} \RB^{\rm ph} e^{+\mathcal{B}_{\rm L}}+ \rho_{\rm R} P_{\rm R} \otimes e^{-\mathcal{B}_{\rm R}} \RB^{\rm ph} e^{+\mathcal{B}_{\rm R}}\nn
&&+\rho_{\rm LR} P_{\rm LR} \otimes e^{-\mathcal{B}_{\rm L}} \RB^{\rm ph} e^{+\mathcal{B}_{\rm R}}\nn
&&+\rho_{\rm RL} P_{\rm RL} \otimes e^{-\mathcal{B}_{\rm R}} \RB^{\rm ph} e^{+\mathcal{B}_{\rm L}}\,.
\end{eqnarray}

%%%%%%%%%%%%%%%%%%%%%%%%%%%%%%%%%%%%%%%%%%%%%%%%%%%%%%%%%%%%%%%%%%%%%%%%%%%%%%%%
%%%%%%%%%%%%%%%%%%%%%%%%%%%%%%%%%%%%%%%%%%%%%%%%%%%%%%%%%%%%%%%%%%%%%%%%%%%%%%%%
%%%%%%%%%%%%%%%%%%%%%%%%%%%%%%%%%%%%%%%%%%%%%%%%%%%%%%%%%%%%%%%%%%%%%%%%%%%%%%%%
%%%%%%%%%%%%%%%%%%%%%%%%%%%%%%%%%%%%%%%%%%%%%%%%%%%%%%%%%%%%%%%%%%%%%%%%%%%%%%%%

\section{Bath correlation functions}\label{App:BCF}

%%%%%%%%%%%%%%%%%%%%%%%%%%%%%%%%%%%%%%%%%%%%%%%%%%%%%%%%%%%%%%%%%%%%%%%%%%%%%%%%
%%%%%%%%%%%%%%%%%%%%%%%%%%%%%%%%%%%%%%%%%%%%%%%%%%%%%%%%%%%%%%%%%%%%%%%%%%%%%%%%

\subsection{Phonon BCF}\label{App:PhononBCF}
We compute the expectation value of the phononic contribution in the Lead-Phonon bath correlation functions, cf. Sec.~\ref{sec:LeadPhononBCF}, given by
\begin{eqnarray}
\mathcal{C}^\sigma_{\rm ph}&=&\expval{e^{-\boldsymbol{\mathcal{B}}_\sigma(\tau)}e^{+\mathcal{B}_\sigma}}\nn
&=&\expval{e^{-\boldsymbol{\mathcal{B}}_\sigma(\tau)+\mathcal{B}_\sigma}}e^{-\com{\boldsymbol{\mathcal{B}}_\sigma(\tau),\mathcal{B}_\sigma}/2}\nn
&=&\expval{e^{\sum_q\left[\frac{h^\ast_{q,\sigma}}{\omega_q}(1-e^{+\ii \omega_q \tau})a^\dagger_q-\frac{h_{q,\sigma}}{\omega_q}(1-e^{-\ii \omega_q \tau})a_q\right]}}\times\nn
&&\times e^{-\ii\sum_q\frac{\abs{h_{q,\sigma}}^2}{\omega_q^2}\sin(\omega_q\tau)}\nn
&=& e^{-\sum_q \abs{\frac{h_{q,\sigma}}{\omega_q}(1-e^{-\ii\omega_q\tau})}^2[n_{\rm B}(\omega_q)+1/2]}\times\nn
&&\times e^{-\ii\sum_q \frac{\abs{h_{q,\sigma}}^2}{\omega_q^2} \sin(\omega_q\tau)}\nn
&=&e^{-\sum_q\frac{\abs{\hqa}^2}{\omega_q^2}[2n_{\rm B}(\omega_q)+1]}\\
&&\times e^{\sum_q\frac{\abs{\hqa}^2}{\omega_q^2}\left\lbrace n_{\rm B}(\omega_q)e^{+\ii\omega_q\tau}+[n_{\rm
B}(\omega_q)+1]e^{-\ii\omega_q\tau}\right\rbrace}\,.\nonumber
\end{eqnarray}
And noting that it is invariant under the transformation $h_{q,\sigma} \to -h_{q,\sigma}$ we conclude
\begin{eqnarray}
\mathcal{C}_{12}(\tau) &=& \mathcal{C}_{\rm ph}^{\rm L}(\tau) \sum_k \abs{t_{k,\rm L}}^2 f_{\rm L}(\upvarepsilon_{k,\rm L}) e^{+\ii\upvarepsilon_{k,\rm L}\tau}\,,\nn
\mathcal{C}_{21}(\tau) &=& \mathcal{C}_{\rm ph}^{\rm L}(\tau) \sum_k \abs{t_{k,\rm L}}^2 [1-f_{\rm L}(\upvarepsilon_{k,\rm L})] e^{-\ii\upvarepsilon_{k,\rm L}\tau}\,,\nn
\mathcal{C}_{34}(\tau) &=& \mathcal{C}_{\rm ph}^{\rm R}(\tau) \sum_k \abs{t_{k,\rm R}}^2 f_{\rm R}(\upvarepsilon_{k,\rm R}) e^{+\ii\upvarepsilon_{k,\rm R}\tau}\,,\nn
\mathcal{C}_{43}(\tau) &=& \mathcal{C}_{\rm ph}^{\rm R}(\tau) \sum_k \abs{t_{k,\rm R}}^2 [1-f_{\rm R}(\upvarepsilon_{k,\rm R})] e^{-\ii\upvarepsilon_{k,\rm R}\tau}\,.
\end{eqnarray}

%%%%%%%%%%%%%%%%%%%%%%%%%%%%%%%%%%%%%%%%%%%%%%%%%%%%%%%%%%%%%%%%%%%%%%%%%%%%%%%%
%%%%%%%%%%%%%%%%%%%%%%%%%%%%%%%%%%%%%%%%%%%%%%%%%%%%%%%%%%%%%%%%%%%%%%%%%%%%%%%%

\subsection{Inter-dot BCF}\label{App:InterdotBCF}
We show explicitly that $\mathcal{C}_{55}(\tau)$ is given by Eq.~(\ref{eq:C55}):
\begin{eqnarray}
\mathcal{C}_{55}(\tau) &=& \expval{e^{\boldsymbol{\mathcal{B}}_{\rm R}(\tau)-\boldsymbol{\mathcal{B}}_{\rm L}(\tau)+\mathcal{B}_{\rm R}-\mathcal{B}_{\rm L}}} e^{+2\ii\Phi}\times\nn
&&\times e^{+[\mathcal{B}_{\rm R}(\tau)-\mathcal{B}_{\rm L}(\tau), \mathcal{B}_{\rm R}-\mathcal{B}_{\rm L}]/2}-\kappa^2\nn
&=& \expval{e^{\sum_q \frac{\lambda_q^*}{\omega_q}(1+e^{+\ii\omega_q\tau}) a_q^\dagger 
-\frac{\lambda_q}{\omega_q}(1+e^{-\ii\omega_q\tau}) a_q}} e^{+2\ii\Phi}\times\nn
&&\times e^{\ii\sum_q \frac{\abs{\lambda_q}^2}{\omega_q^2} \sin(\omega_q\tau)}-\kappa^2\nn
&=& e^{+2\ii\Phi} e^{-\sum_q \frac{\abs{\lambda_q}^2}{\omega_q^2}\left[\left(1+n_{\rm B}(\omega_q)\right)e^{-\ii\omega_q\tau} 
+ n_{\rm B}(\omega_q) e^{+\ii\omega_q\tau}\right]}\times\nn
&&\times e^{-\sum_q \frac{\abs{\lambda_q}^2}{\omega_q^2}\left(1+2n_{\rm B}(\omega_q)\right)}-\kappa^2\\
&=& \kappa^2 \left[e^{-\sum_q \frac{\abs{\lambda_q}^2}{\omega_q^2}\left[\left(1+n_{\rm B}(\omega_q)\right)e^{-\ii\omega_q\tau} 
+ n_{\rm B}(\omega_q) e^{+\ii\omega_q\tau}\right]}-1\right]\,,\nonumber
\end{eqnarray}
where $\lambda_q=h_{qL}-h_{qR}$.
The bath correction function $\mathcal{C}_{66}(\tau)$ can be obtained via $\mathcal{C}_{66}(\tau) \hat= \mathcal{C}_{55}^*(-\tau)$.
We show explicitly that $\mathcal{C}_{56}(\tau)$ is given by Eq.~(\ref{eq:C56}):
\begin{eqnarray}
\mathcal{C}_{56}(\tau) &=& \expval{e^{\boldsymbol{\mathcal{B}}_{\rm R}(\tau)-\boldsymbol{\mathcal{B}}_{\rm L}(\tau)-(\mathcal{B}_{\rm R}-\mathcal{B}_{\rm L})}}\times\nn
&&\times e^{-[\boldsymbol{\mathcal{B}}_{\rm R}(\tau)-\boldsymbol{\mathcal{B}}_{\rm L}(\tau), \mathcal{B}_{\rm R}-\mathcal{B}_{\rm L}]/2} - \abs{\kappa}^2\nn
&=& \expval{e^{\sum_q \frac{\lambda_q^*}{\omega_q}(e^{+\ii\omega_q\tau}-1) a_q^\dagger
-\frac{\lambda_q}{\omega_q}(e^{-\ii\omega_q\tau}-1)a_q}}\times\nn
&&\times e^{-\ii\sum_q\frac{\abs{\lambda_q}^2}{\omega_q^2} \sin(\omega_q\tau)} - \abs{\kappa}^2\nn
&=& \abs{\kappa}^2\times\\
&&\times \left[e^{+\sum_q \frac{\abs{\lambda_q}^2}{\omega_q^2}\left[\left(1+n_{\rm B}(\omega_q)\right)e^{-\ii\omega_q\tau} 
+ n_{\rm B}(\omega_q) e^{+\ii\omega_q\tau}\right]}-1\right]\,.\nonumber
\end{eqnarray}
The bath correction function $\mathcal{C}_{65}(\tau)$ can be obtained via the KMS-condition yielding $\mathcal{C}_{56}(\tau) \hat= \mathcal{C}_{65}(\tau)$.

%%%%%%%%%%%%%%%%%%%%%%%%%%%%%%%%%%%%%%%%%%%%%%%%%%%%%%%%%%%%%%%%%%%%%%%%%%%%%%%%
%%%%%%%%%%%%%%%%%%%%%%%%%%%%%%%%%%%%%%%%%%%%%%%%%%%%%%%%%%%%%%%%%%%%%%%%%%%%%%%%
%%%%%%%%%%%%%%%%%%%%%%%%%%%%%%%%%%%%%%%%%%%%%%%%%%%%%%%%%%%%%%%%%%%%%%%%%%%%%%%%
%%%%%%%%%%%%%%%%%%%%%%%%%%%%%%%%%%%%%%%%%%%%%%%%%%%%%%%%%%%%%%%%%%%%%%%%%%%%%%%%

\section{Symmetries in the Characteristic polynomials}\label{App:symmetries}

To show these symmetries, we show separate symmetries of the terms in the characteristic polynomial:

First, we note that trivially, the combination $\mathcal{L}_{23} \mathcal{L}_{32}$ does not depend on
counting fields and is thus by construction inert to symmetry transformations of type~(\ref{eq:symmetry}).

\begin{widetext}
Second, one can directly show that terms of the form
$\mathcal{L}_{12}\mathcal{L}_{21}$, $\mathcal{L}_{13}\mathcal{L}_{31}$, $\mathcal{L}_{24}\mathcal{L}_{42}$, and $\mathcal{L}_{34}\mathcal{L}_{43}$
are also invariant under such transformations.
We only show this explicitly for the first combination (the proof is analogous for the other terms), where we have
\begin{eqnarray}
\mathcal{L}_{12} &=& \sum_{\f{n}} \left(\gL^{0-,-\f{n}} e^{-\ii\chi} e^{-\ii\xi(\eM-\eO+\f{n}\cdot\f{\Omega})} e^{+\ii\phi\f{n}\cdot\f{\Omega}} 
+ \gR^{0-,-\f{n}} e^{+\ii\phi\f{n}\cdot\f{\Omega}}\right)\,,\nn
\mathcal{L}_{21} &=& \sum_{\f{n}} \left(\gL^{-0,+\f{n}} e^{+\ii\chi} e^{+\ii\xi(\eM-\eO+\f{n}\cdot\f{\Omega})} e^{-\ii\phi\f{n}\cdot\f{\Omega}} 
+ \gR^{-0,+\f{n}} e^{-\ii\phi\f{n}\cdot\f{\Omega}}\right)\,.
\end{eqnarray}
We can use the detailed balance relations~(\ref{eq:local_detailed_balance}) to rewrite e.g.\  the first matrix element as (now keeping the counting fields explicitly)
\begin{eqnarray}
\mathcal{L}_{12}(\chi,\xi,\phi) &=& \sum_{\f{n}} \Big(
\gL^{-0,+\f{n}} e^{-\ii\chi} e^{-\ii\xi(\eM-\eO+\f{n}\cdot\f{\Omega})} e^{+\ii\phi\f{n}\cdot\f{\Omega}} e^{+\bL(\eM-\eO-\mL+\f{n}\cdot\f{\Omega})} e^{-\beta_{\rm ph} \f{n}\cdot\f{\Omega}}\nn
&&+ \gR^{-0,+\f{n}} e^{+\bR(\eM-\eO-\mR+\f{n}\cdot\f{\Omega})} e^{-\beta_{\rm ph} \f{n}\cdot\f{\Omega}}\Big)\nn
&=& e^{\bR(\eM-\eO-\mu_R)} \mathcal{L}_{21}(-\chi+\ii(\bL\mL-\bR\mR), -\xi+\ii(\bR-\bL), -\phi+\ii(\bR-\beta_{\rm ph}))\,.
\end{eqnarray}
\end{widetext}
With the short-hand notation
$\mathcal{L}_{ij}^- = \mathcal{L}_{ij}(-\f{\chi})$ and 
$\bar{\mathcal{L}}_{ij} = \mathcal{L}_{ij}(\f{\chi}+\ii\f{\Delta})$ where
$\f{\Delta}=(\bL\mL-\bR\mR, \bR-\bL, \bR-\beta_{\rm ph})$
we can summarize the relations
\begin{eqnarray}\label{EQ:symmetry1}
\mathcal{L}_{12}^- &=& e^{+\bR(\eM-\eO-\mR)} \bar{\mathcal{L}}_{21}\,,\nn
\mathcal{L}_{21}^- &=& e^{-\bR(\eM-\eO-\mR)} \bar{\mathcal{L}}_{12}\,,\nn
\mathcal{L}_{13}^- &=& e^{+\bR(\eP-\eO-\mR)} \bar{\mathcal{L}}_{31}\,,\nn
\mathcal{L}_{31}^- &=& e^{-\bR(\eP-\eO-\mR)} \bar{\mathcal{L}}_{13}\,,\nn
\mathcal{L}_{24}^- &=& e^{+\bR(\ePM-\eM-\mR)} \bar{\mathcal{L}}_{42}\,,\nn
\mathcal{L}_{42}^- &=& e^{-\bR(\ePM-\eM-\mR)} \bar{\mathcal{L}}_{24}\,,\nn
\mathcal{L}_{34}^- &=& e^{+\bR(\ePM-\eP-\mR)} \bar{\mathcal{L}}_{43}\,,\nn
\mathcal{L}_{43}^- &=& e^{-\bR(\ePM-\eP-\mR)} \bar{\mathcal{L}}_{34}\,,
\end{eqnarray}
such that e.g.\  products of the form $\mathcal{L}_{12}\mathcal{L}_{21}$ are invariant under the transformations~(\ref{eq:symmetry}), 
i.e., $\mathcal{L}_{12}^-\mathcal{L}_{21}^-=\bar{\mathcal{L}}_{12}\bar{\mathcal{L}}_{21}$.

Third, we consider combinations of three off-diagonal matrix elements by noting the additional symmetry
\begin{eqnarray}\label{EQ:symmetry2}
\mathcal{L}_{23}^- &=& e^{+\bR(\eP-\eM)} \bar{\mathcal{L}}_{32}\,,\nn
\mathcal{L}_{32}^- &=& e^{-\bR(\eP-\eM)} \bar{\mathcal{L}}_{23}\,,
\end{eqnarray}
which together with the symmetries in Eq.~(\ref{EQ:symmetry1}) can be used to show that in 
the characteristic polynomial~(\ref{eq:charPol})
the terms with three off-diagonal matrix elements
are also inert under the transformations~(\ref{eq:symmetry}), i.e.,
\begin{eqnarray}
\mathcal{L}_{23}^-\mathcal{L}_{34}^-\mathcal{L}_{42}^-+\mathcal{L}_{24}^-\mathcal{L}_{43}^-\mathcal{L}_{32}^-
&=&\bar{\mathcal{L}}_{23}\bar{\mathcal{L}}_{34}\bar{\mathcal{L}}_{42}
+\bar{\mathcal{L}}_{24}\bar{\mathcal{L}}_{43}\bar{\mathcal{L}}_{32}\,,\nn
\mathcal{L}_{12}^-\mathcal{L}_{23}^-\mathcal{L}_{31}^-+\mathcal{L}_{13}^-\mathcal{L}_{32}^-\mathcal{L}_{21}^-
&=&
\bar{\mathcal{L}}_{12}\bar{\mathcal{L}}_{23}\bar{\mathcal{L}}_{31}+\bar{\mathcal{L}}_{13}\bar{\mathcal{L}}_{32}\bar{\mathcal{L}}_{21}\,.\nn
\end{eqnarray}

Finally, we note that the terms $\mathcal{L}_{12}\mathcal{L}_{21}\mathcal{L}_{34}\mathcal{L}_{43}$ and $\mathcal{L}_{13}\mathcal{L}_{31}\mathcal{L}_{24}\mathcal{L}_{42}$ 
can be treated similarly to the terms with just two off-diagonal matrix elements, and that the last two terms in the characteristic polynomial~(\ref{eq:charPol})
obey
\begin{eqnarray}
\mathcal{L}_{12}^-\mathcal{L}_{24}^-\mathcal{L}_{43}^-\mathcal{L}_{31}^-+\mathcal{L}_{13}^-\mathcal{L}_{34}^-\mathcal{L}_{42}^-\mathcal{L}_{21}^-\nn
=
\bar{\mathcal{L}}_{12}\bar{\mathcal{L}}_{24}\bar{\mathcal{L}}_{43}\bar{\mathcal{L}}_{31}+\bar{\mathcal{L}}_{13}\bar{\mathcal{L}}_{34}\bar{\mathcal{L}}_{42}\bar{\mathcal{L}}_{21}
\end{eqnarray}
which can be shown with Eqns.~(\ref{EQ:symmetry1}).

\end{document}